\documentclass[10pt,twocolumn,letterpaper]{article}
\pdfoutput=1 
\usepackage{iccv}
\usepackage{times}
\usepackage{epsfig}
\usepackage{graphicx}
\usepackage{amsmath}
\usepackage{amssymb}
\usepackage{subfigure}
\usepackage{graphicx}
 
\usepackage{subfigure}
\usepackage{dblfloatfix}
\usepackage{xcolor}
\usepackage{caption}
\usepackage{amsmath,enumitem}
\usepackage{float}
\usepackage{contour}
\usepackage{booktabs}
\usepackage{multirow}
\usepackage{authblk}

\usepackage[title]{appendix}

\usepackage{hyperref}
\hypersetup{breaklinks=true,bookmarks=false}

\definecolor{cite_color}{RGB}{0,200,0}
\hypersetup{
    colorlinks=true,
    linkcolor=blue,
    filecolor=magenta,      
    urlcolor=black,
    citecolor=cite_color,
    pdftitle={Drones4Good: Supporting Disaster Relief Through Remote Sensing and AI},
    pdfpagemode=FullScreen,
    }

\iccvfinalcopy 

\def\iccvPaperID{22} 

\ificcvfinal\fi

\begin{document}


\title{Drones4Good: Supporting Disaster Relief Through Remote Sensing and AI}

\author[1]{Nina Merkle}
\author[1]{Reza Bahmanyar}
\author[1]{Corentin Henry}
\author[1]{Seyed Majid Azimi}
\author[1]{Xiangtian Yuan}
\author[2]{Simon Schopferer}
\author[1]{Veronika Gstaiger}
\author[1]{Stefan Auer}
\author[3]{Anne Schneibel}
\author[3]{Marc Wieland}
\author[4]{Thomas Kraft}

\affil[ ]{German Aerospace Center (DLR)} 
\affil[ ]{$^{1}$\text{Remote Sensing Technology Institute},\ \  $^{2}$\text{Institute of Flight Systems}} 
\affil[ ]{$^{3}$\text{Remote Sensing Data Center},\ \ $^{4}$\text{Institute of Optical Sensor Systems}}
\affil[ ]{\tt\small contact: firstname.lastname@dlr.de}

\maketitle
\ificcvfinal\thispagestyle{empty}\fi

\begin{abstract}

In order to respond effectively in the aftermath of a disaster, emergency services and relief organizations rely on timely and accurate information about the affected areas. Remote sensing has the potential to significantly reduce the time and effort required to collect such information by enabling a rapid survey of large areas. To achieve this, the main challenge is the automatic extraction of relevant information from remotely sensed data. In this work, we show how the combination of drone-based data with deep learning methods enables automated and large-scale situation assessment. In addition, we demonstrate the integration of onboard image processing techniques for the deployment of autonomous drone-based aid delivery. The results show the feasibility of a rapid and large-scale image analysis in the field, and that onboard image processing can increase the safety of drone-based aid deliveries. 

\end{abstract}

\vspace{-1.5mm}
\section{Introduction}

Every year, millions of people around the world are affected by natural and man-made disasters~\cite{CRED2023}. In order to respond effectively to such crises, emergency services and relief organizations rely on timely, comprehensive, and accurate information about the disaster's extent. For years, emergency mapping has been based on remote sensing data to support rescue operations, gathering information on affected areas by comparing images acquired before and after the event by satellites, aircrafts, or drones~\cite{Lang2019}. However, the automatic extraction of such information and its rapid, scalable, and reliable delivery is still a challenge. Recent developments in computer vision and the rapid evolution of graphics processing units have led to optimized, fast-running algorithms, opening up new possibilities in disaster and humanitarian relief~\cite{Beduschi20}.

\begin{figure}[t] 
	\centering
{\includegraphics[width=0.445\textwidth]{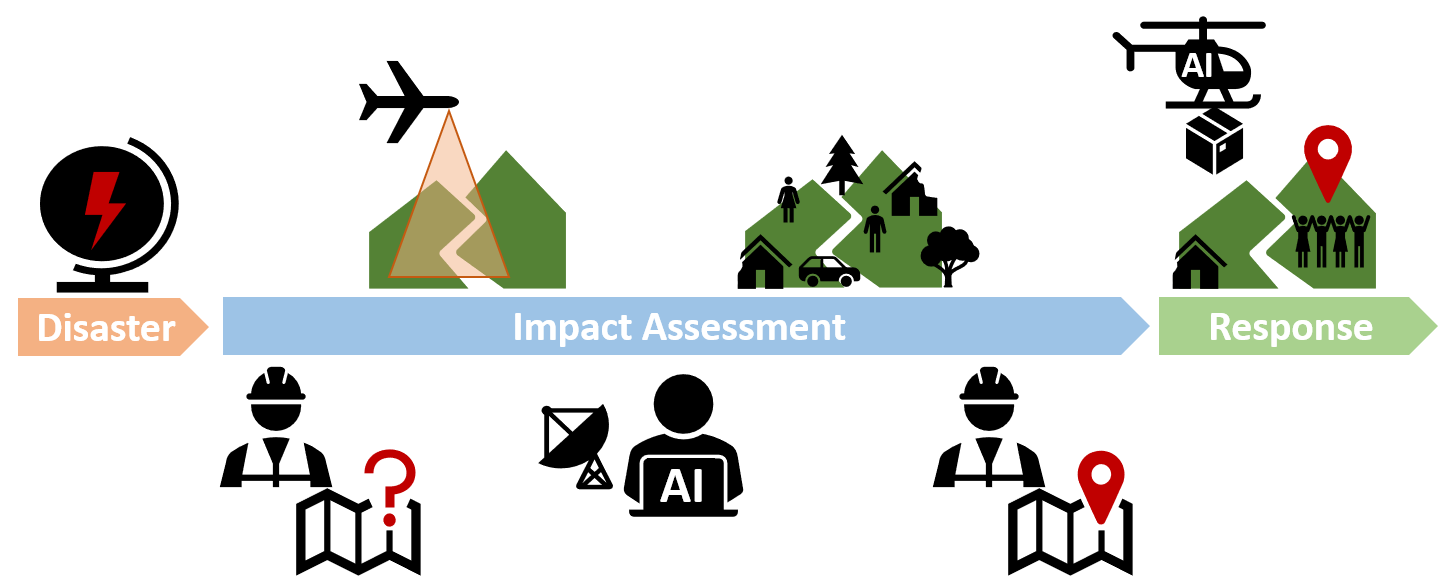}}
    \vspace{-2.5mm}
	\caption{Illustration of the presented workflow.} 
	\label{fig:workflow}
\end{figure}

In this paper, we explore the potential of state-of-the-art deep learning techniques for image analysis in combination with remote sensing data acquired by drones. The workflow, ranging from real-time and large-scale image acquisition and mapping, to automatic and fast image analysis, over to an onboard image processing method to support automatic aid delivery by drones is presented in \autoref{fig:workflow}. The extracted information can be seamlessly integrated into an organization's operations control center to effectively support emergency services and aid organizations in their operations. The methodological focus is on: (1) automated, near real-time extraction of roads, buildings and people for initial impact assessment to help prepare missions in terms of logistics and routing of relief forces and supplies, and (2) real-time detection of people from drones to increase the safety of drone-based aid delivery once target areas have been identified. Here we use the term "near real-time" to refer to processing within hours, and "real-time" to refer to processing within seconds. The performance of all methods has been thoroughly tested and evaluated, demonstrating their potential to rapidly analyze large amounts of data and increase the safety of drone-based aid deliveries. 

Ethically, there is no recognition of individuals and no sensitive data stored in a cloud where privacy could be compromised. Our work has been carried out in close collaboration with organizations such as the World Food Programme (WFP), I.S.A.R. Germany, and the Bavarian Red Cross (BRK), who recognize the high potential of our methods and the urgent need to put them into practice.

\vspace{-1.5mm}
\section{Situation Assessment}\label{sec:situation_assessment}

In order to support relief efforts in rapidly assessing the impact of a disaster, we present an approach consisting of two-steps: 1) mapping the scene in real time with a camera system mounted on a drone in order to provide up-to-date image data over the area of interest, and 2) automatically extracting relevant information in the field and in near-real time to help humanitarian organizations assess the acquired data faster. For the mapping of the disaster area, the rapid mapping camera system MACS-Micro~\cite{Hein2019} is used, which is carried and integrated into a fast-flying drone. The image data are available on the ground in real time using a commercial radio link. If the range of the radio link is exceeded, the data will be available immediately after landing. The system consists of nadir-pointing cameras, a GNSS receiver combined with an industrial-grade inertial measurement unit, an embedded computing unit, and a radio link. 
During a typical campaign, the camera is operated at an altitude of $200$m above ground, with a speed of $80$km/h, and a frame rate of 2Hz. This gives an acquisition rate of around $3200$m$^2$ per second with a GSD of $3$cm. The resulting product is a scaled image mosaic showing the current situation of the disaster area, which can be used as an additional map layer for common geographic information systems.


According to feedback from relief organizations involved, a number of important questions arise in the first moments after a disaster, such as: which areas are most affected? How many people have been affected and where are they now? Which infrastructure can still be used for rescue and relief?  Based on these questions, we identified the objects "roads, "buildings" and "people" as the most important in helping emergency responders to give timely answers to these questions. To this end, we developed three algorithms running on a GPU laptop in the field.

\textbf{Road segmentation:} We use a Dense-U-Net-121~\cite{Henry_2020_Aerial_Noise} based on the widely-used U-Net~\cite{Ronneberger_2015_UNet_Segmentation}. For the backbone of both the encoder and decoder, we use a DenseNet-121 as it offers the best compromise between the accuracy of the result and the computational resources required. The resulting road mask provides rescue teams with an up-to-date map of the captured area and can be used to identify cut-off regions. In addition, changes to the road network and severely affected areas can be quickly identified when compared to a pre-disaster scene.

\textbf{Building segmentation:} For the segmenting of buildings, the HRNet~\cite{wang2020deep} consisting of four parallel, multi-resolution streams that maintain fine-grained features throughout the network is used. This feature allows for more precise localization, which is crucial for the task of building segmentation. The resulting building mask can be used to identify populated areas and, if compared to a pre-disaster scene, to estimate the number of people affected and damaged houses.

\textbf{Person detection:} For the detection of people, we use an adapted YOLOv3~\cite{redmon2018yolov3} object detection method that addresses challenges such as variations in scenes, poses, scales, and viewing angles posed by images during real humanitarian missions. As no publicly available person detection dataset was suitable for our case, we created a new dataset consisting of aerial and drone images covering different scenarios and  countries (for more details see \cite{Bahmanyar2023personDetection}). The output of the model is the location of each detected person in the form of bounding boxes, which is extremely valuable information for search and rescue missions or for the safe delivery of supplies to affected areas.

\begin{figure}[t]
	\centering
        \subfigure{\includegraphics[width=0.29\textwidth]{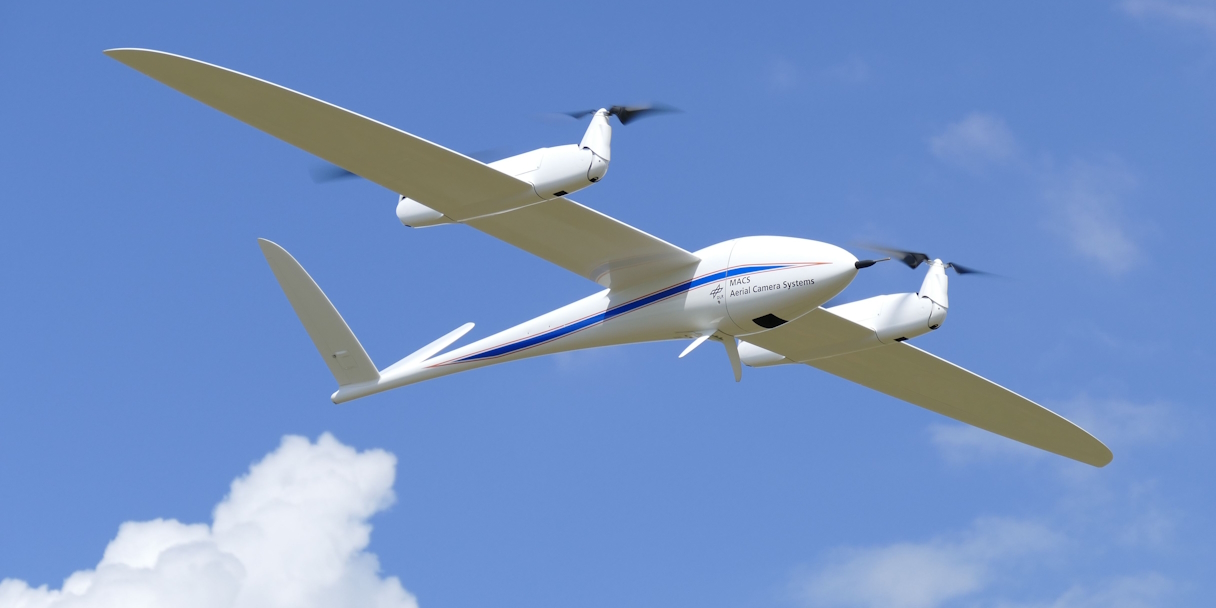}}
        \subfigure{\includegraphics[width=0.104\textwidth]{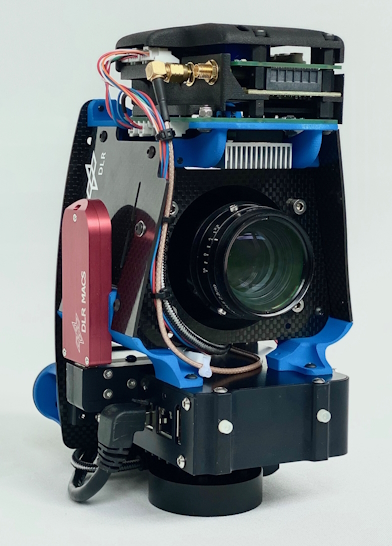}}
        \vspace{-3.5mm}
	\caption{Vertical take-off and landing drone (left) with integrated camera system MACS-Micro (right).}
	\label{fig:MACS}
\end{figure}

After the inference phase, the models' predictions are assigned the same geo-referenced coordinate space as the input image. The output layers are saved as a GeoTIFF file and overlaid on the geo-referenced input image for further analysis in any software supporting this format. The acquired image data as well as the derived information layers can either be shared directly with interested humanitarian organizations or delivered to institutions such as the Center for Satellite Based Crisis Information ZKI~\cite{ZKI23}, where earth observation data are analyzed and situational awareness is generated before, during and after a natural or man-made disaster in form of ISO standardized products.

\vspace{-1.5mm}
\section{Delivery of relief supplies}

\begin{figure}[t!] 
	\centering
{\includegraphics[width=0.30\textwidth]{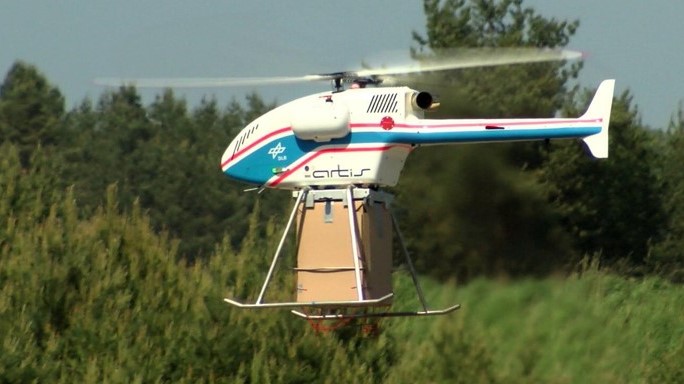}}
    \vspace{-2.5mm}
	\caption{Unmanned helicopter superARTIS equipped with payload for aerial delivery.} 
	\label{fig:drone}
\end{figure}

After assessing the impact of a disaster, the next phase is to response. We focus on a specific case where we assume that people are cut off and can only be reached by air. Relief supplies are delivered by a drone dropping a payload. To increase the safety of people on the ground during the process, we investigated how a camera system can be combined with an AI algorithm to detect people onboard the drone.  Generally, various drone configurations could be envisioned to deliver goods in humanitarian aid scenarios. In this work, we focused on a {superARTIS} demonstrator platform which is equipped with a box drop mechanism as depicted in \autoref{fig:drone}. This box-dropping payload was integrated and flight tested in cooperation with Wings For Aid~\cite{Dauer2022}. For assessing the safety of the drop zone, the drone is also equipped with a downward looking camera and an onboard processing unit. 
The capability of running AI-based people detection onboard in real time may serve different purposes depending on the automation desired for the operation and the availability of a data link. In case the aircraft is remotely piloted via a low bandwidth data link, the processing results can still be transmitted with a low bandwidth demand as opposed to transmitting the video stream. This enables the remote pilot to assess whether or not it is safe to release the box of supplies. Also, no personal data is transmitted or recorded in this scenario. When no data link is available, the onboard autopilot may use the onboard person detection to conduct the delivery autonomously in a safe manner.

In order to enable real-time person detection on drones, we optimize the processing of the YOLOv3 discussed in \autoref{sec:situation_assessment}. We reduce the data type precision to float16 and simplify the non-maximum suppression procedure which is one of the most computationally intensive steps. 
The images acquired by the camera system are transferred to and processed by the onboard GPU. The detection results are then passed to the CPU and geolocated. We use the cross-platform data format Protocol Buffers~\cite{Varda2008} to ensure efficient data transfer rates and communication within the system in order to minimize the overall processing time.

\section{Experiments}\label{sec:results}

\paragraph{Training parameters:}

For \textbf{road segmentation}, a Dense-U-Net-121 was trained on the DeepGlobe18~\cite{Demir2018DeepGlobe2A} dataset consisting of 1,632km$^2$ annotated images data at 50cm/px from southeast Asian regions. We trained it for 40 epochs with a patch size of $512\times512$px and a batch size of 12. For \textbf{building segmentation}, we used the Inria dataset~\cite{INRIA_dataset} to train the HRNet. This dataset contains 405~km$^2$ of labeled image data at 20cm/px from the USA and Austria. The training was performed for 20 epochs with a patch size of $512\times512$ and a batch size of 16. For training the \textbf{person detection} network, we used our own training dataset~\cite{Bahmanyar2023personDetection} consisting of 10,050 annotated persons in 311 aerial and drone images (train: 259, validation: 25, test: 27), with GSDs ranging from 0.2 to 6cm/pixel, covering areas in Germany, the Netherlands, Switzerland, Spain, France, and Nepal. 

\paragraph{Test hardware:} To use our models in humanitarian contexts, they must run on portable and affordable computers. Therefore, we chose an Alienware Area51m laptop with 32GB of RAM and an NVIDIA RTX 2080 Super with 8GB of VRAM to process the image patches. The processing time of the three models is summarized in \autoref{table:laptop_inference}. For the onboard processing, we used a Jetson AGX Xavier with 8GB of VRAM. Here we tiled each image into patches of 416$\times$416px with 10\% overlap to fit into the GPU memory.

\begin{figure}[t]
    \centering
    \includegraphics[width=0.75\columnwidth]{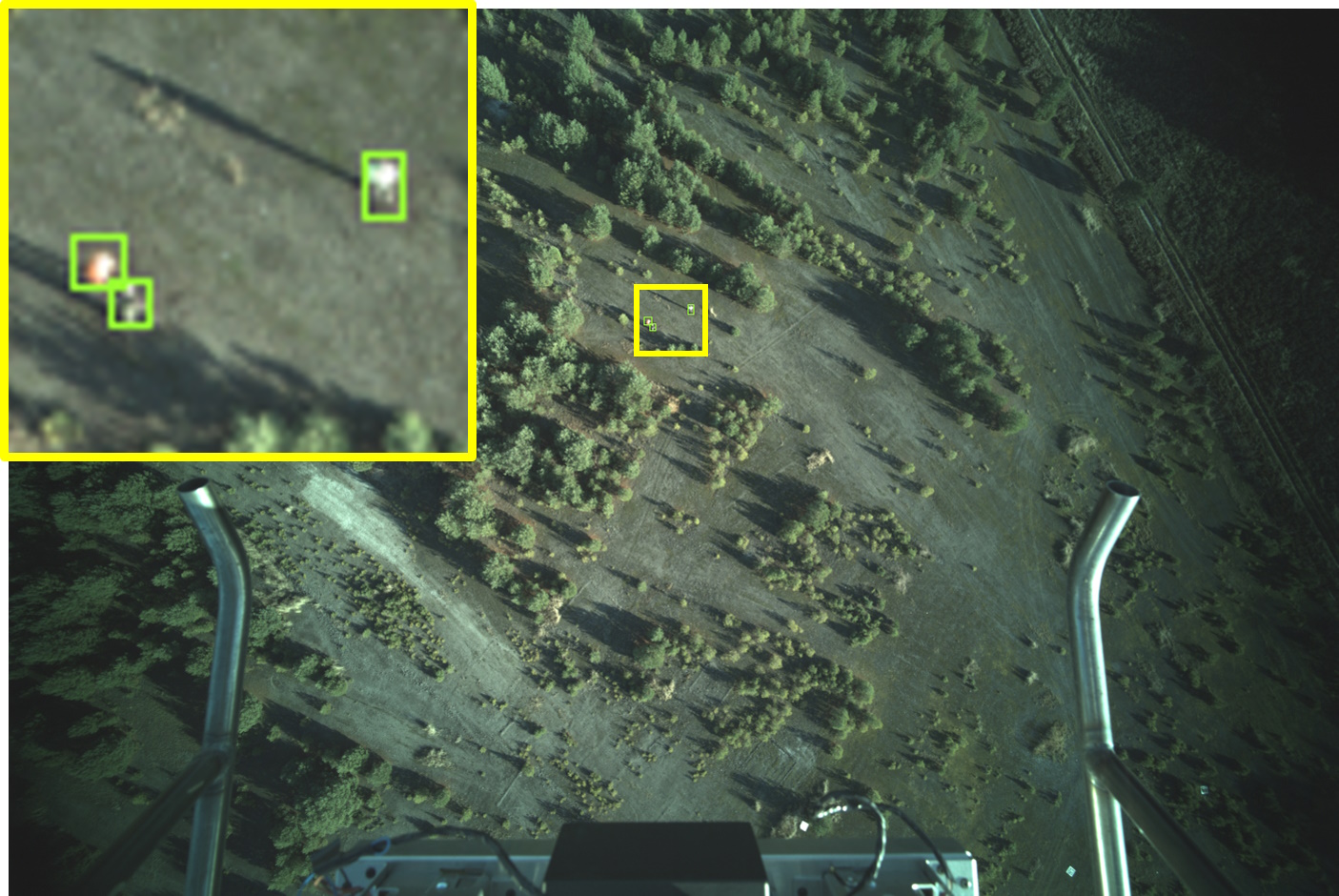}
         \vspace{-1.5mm}
    \caption{View from the delivery drone and demonstration of the onboard person detection algorithm.}
    \label{fig:resut_person_detection_onboard}
\end{figure}

\begin{figure*}[t!] 
	\centering
{\includegraphics[width=0.86\textwidth]{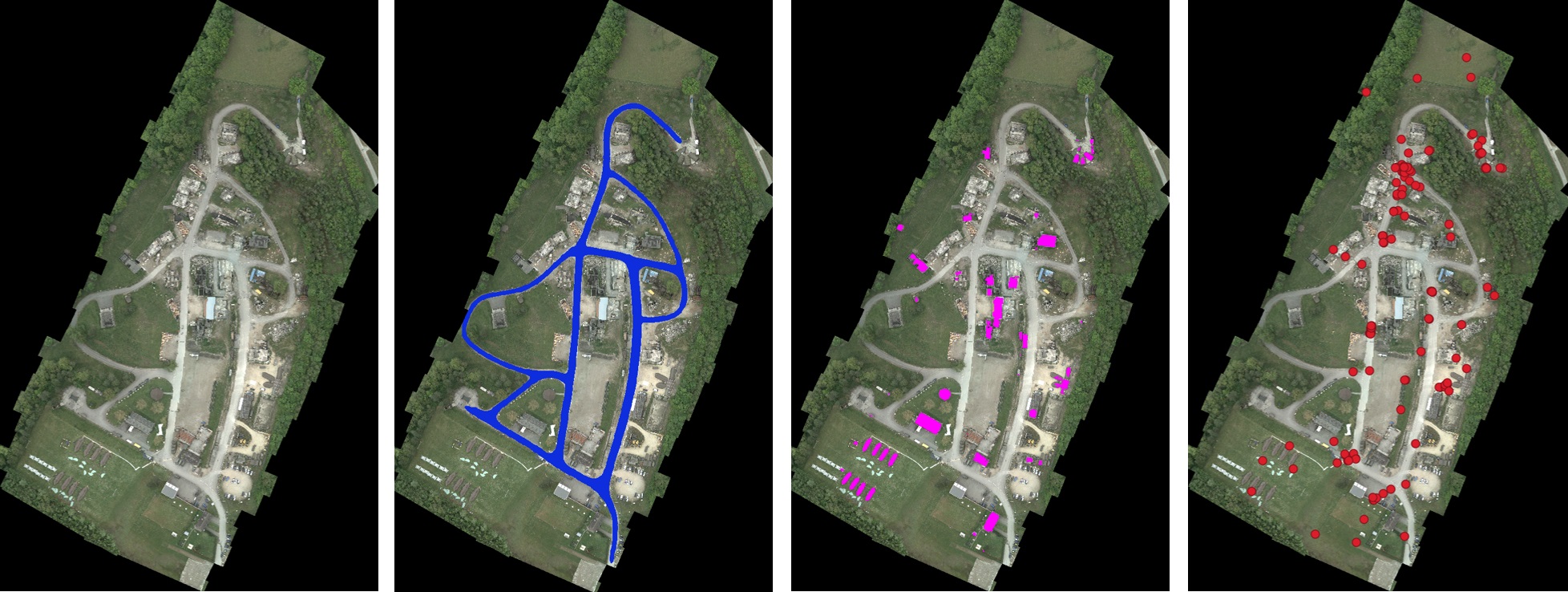}}
     \vspace{-2.5mm}
	\caption{Qualitative results: image mosaic, segmented roads, segmented buildings, and detected people (left to right).} 
	\label{fig:results_near_real-time}
\end{figure*}

\begin{table*}[t]
\caption{Processing speed (per megapixel and per area) and quantitative results obtained on an Alienware Area51m laptop. }\label{table:laptop_inference}
\vspace{-3.5mm}
\resizebox{\textwidth}{!}{
\begin{tabular}{@{\hspace{0pt}}lccccccccccc@{}}
\toprule
\textbf{\multirow{2}{*}{Task}}    & \textbf{Training} &     \textbf{GSD} & \multicolumn{2}{c}{\textbf{Computational time}} &\textbf{Complet.} & \textbf{Correct.} & \textbf{Quality} & \textbf{Prec. } & \textbf{Recall}  & \textbf{ IoU }&  \textbf{AP} \\
     & \textbf{data}      & \textit{[cm]} & \textit{per MP} & \textit{per  km$^2$} & \textit{[\%]} & \textit{[\%]}  & \textit{[\%]}  & \textit{[\%]}  & \textit{[\%]}  & \textit{[\%]} &  \textit{[\%]}  \\
\midrule
Road             &  DeepGlobe18~\cite{Demir2018DeepGlobe2A}           & 50       & 0.80s            & 3.30s & 70.96 & 76.48  & 58.08  & - & - & - & -  \\
\midrule
\multirow{1}{*}{Building } & Inria~\cite{INRIA_dataset} & 20      & 0.38s              & 9.50s    &- & -&- & 
83.74 & 77.70 & 68.12 & -  \\
\midrule
\multirow{1}{*}{People }  &  Ours~\cite{Bahmanyar2023personDetection}  & 3       & 0.44s               & 19min &- &- &- & 54.13 & 65.87 &- &60.36  \\
                                  
\bottomrule
\end{tabular}}
\end{table*}


\vspace{-5.5mm}

\paragraph{Results \& Discussion:}

When applying research methods to real-life applications, the overall framework and all its requirements must be carefully considered and taken into account during the development and implementation of algorithms. For our specific application case, some of the constraints are due to technical, computational, legal, and financial limitations, while others are due to the overall situation during a disaster. But also the algorithms used come with limitations. For our models to generalize well to new locations, the images must be as similar as possible to those of our training set. Ideally, they should be taken at nadir, i.e. not be side-looking, in clear weather, and with sufficient illumination. The resolution of the images acquired after the disaster might be higher than during training, in which case the images are downsampled if necessary.

\textbf{For the road segmentation task}, we selected 20 test scenes to evaluate the generalization capability of our model, each annotated by hand with vector lines: 1 from Epeisses, Switzerland in a disaster training area (0.2km$^2$), 10 from the Ahr Valley, Germany after the major flooding event in 2021 (10.5km$^2$), and 9 from Beira, Mozambique after the Cyclone Idai in 2019 (3.5km$^2$). While our model was already shown to perform well in many regions around the world~\cite{Henry_2020_Aerial_Noise}, e.g. in Nepal (see appendix), it also achieved excellent results in the chosen test scenes as in~\autoref{fig:results_near_real-time}. Most roads were successfully detected with a completeness of 71\%, and few false positives with a correctness of 76\% (metrics from~\cite{Wiedemann_1998_Empirical_Axes}, cf.~\autoref{table:laptop_inference}). The predicted roads are regular and continuous despite changes in color and material. Some sections were incorrectly  detected for three reasons: 1) the model was trained to detect roads, but not larger asphalted areas such as parking areas or certain dead ends, 2) after disasters, sand, mud, and debris may be present on the road, which our training dataset does not feature, and 3) drone image mosaics may contain irregular regions of background along their borders, which removes the necessary context to correctly identify short sections of roads. Despite these obstacles, our model achieved 58\% quality. 

\vspace{-0.5mm}
\textbf{For the building segmentation task}, we used the same 11 scenes as for the road segmentation from Epeisses and the Ahr Valley, and 3 separate scenes from Beira (2.8km$^2$) for evaluation with manually annotated ground truth. The average precision, recall, and intersection over union (IoU) scores are reported in \autoref{table:laptop_inference}). In contrast to the Ahr Vally scene, the numbers for Epeisses and Beira are much lower. There are three main reasons why: 1) the data has a  much higher resolution than the training data and therefore has different spectral and textural features, 2) in Beira, a large number of buildings are very small, which is not reflected in the training data, and 3) in Epeisses, the scene includes large tents that are mistakenly segmented as buildings.

\textbf{For the people detection task}, we evaluate the trained model on a test set of 27 images with 410 annotations. One of the challenges we faced was the altitude required for the delivery drone. In order to safely drop the supplies, the flight altitude must be around 80m, resulting in images with a ground sampling distance of $1$-$3$cm. Therefore, the training set for the person detection algorithm had to be adjusted to include more images in this GSD range. In addition, images with a GSD of less than $6$cm had to be removed from the training set, as the visual appearance of people varies too much between $1$-$10$cm. Overall, we achieved a precision, recall, and average precision (AP) of around 54\%, 66\%, and 60\%, respectively (see \autoref{table:laptop_inference}). Images with a lower GSD generally give better predictions, although the model struggles with complex backgrounds such as vegetation or disaster ruins, and is affected by changes in camera angle. Increasing the variety of the training data could help to overcome these limitations. Testing the optimized model on the Jetson board achieved a 2s processing time for each 16MP image without compromising accuracy and recall. \autoref{fig:resut_person_detection_onboard} shows a sample result with a GSD of $3$cm/pixel.

\vspace{-1.5mm}
\section{Conclusion \& Future Work}

Our results show that the combination of computer vision and remote sensing technologies has great potential to significantly improve disaster management and humanitarian aid. Fast and large-scale image analysis becomes feasible, and onboard image processing can increase the safety of drone-based aid deliveries. However, to improve their generalization and performance, these methods need to be trained on larger datasets from around the world, tested in the field, and extended to include other relevant features such as road or building damage. End-user feedback and knowledge will play an important role in the future development and improvement of the technologies. On the other hand, limitations in the algorithms should be overcome to make the models more robust to changes in the images (e.g. viewing angle) and trainable with less labeled data. %
{\small
\bibliographystyle{ieee_fullname}
\bibliography{egbib}
}

\newpage

\begin{appendices}
\section{Supplementary Material}

\subsection{Training parameters}
For \textbf{road segmentation}, our Dense-U-Net-121 was trained using an NVIDIA RTX Titan GPU, with ImageNet pre-training for the encoder, a cross-entropy loss, an ADAM optimizer, an initial learning rate of $1e-4$ and an exponential learning rate decay of 0.8 applied after each epoch. We applied random horizontal flips and $90^\circ$ rotations. A quantile truncation and a normalization were applied on each input channel separately to remove the upper and lower 2\% of outliers.

For \textbf{building segmentation}, ImageNet weights were used to initialize the model. The training was performed on four NVIDIA RTX Titan GPUs using a cross entropy loss with online hard example mining~\cite{shrivastava2016training}, stochastic gradient descent (SGD) optimizer with an initial learning rate of 0.01, weight decay of 0.001, and Nestrov momentum of 0.9. The learning rate was decayed and the training patches are randomly flipped and rescaled. The image processing is the same as for the road segmentation.

For training the \textbf{person detection} network we use a batch size of 50 and adopted a learning rate of $5\mathrm{e}{-7}$, using the scheduling mechanism from \cite{redmon2018yolov3}. Adam optimizer with a decay of $5\mathrm{e}{-4}$ and a pre-trained model from the MS~COCO dataset~\cite{COCO} was used. We performed a statistical analysis for selecting the anchor boxes. To calculate the loss, we used a combination of logistic regression for objectness error and complete IoU~\cite{Zheng2022} for bounding box error, similar to~\cite{redmon2018yolov3}.

\subsection{Details about the test data for road and building segmentation}

Besides the MACS camera system~\cite{Hein2019}, image data from various systems were used. For these scenes, the data was acquired by the 4K camera system~\cite{Kurz2014} and from Germany's Federal Agency for Cartography and Geodesy (Digital Orthophotos, DOP20). All test images are resampled to 50~cm/pixel for the road segmentation model and to 20~cm/pixel for the building segmentation model. In addition, in order to test the generalizability of the models in different regions of operation, particularly in disaster-prone developing countries, the trained models are tested in Beira, Mozambique, and Kathmandu, Nepal, following cyclones and earthquakes. The UAV image of Beira was provided by WFP and Mozambique's National Institute for Disaster Management (INGC), and the aerial imagery of Kathmandu was captured by the MACS system.

\subsection{Detailed results for road segmentation}

\begin{figure*}[h!]
\centering 
    \subfigure[The complete scene in Epeisses with overlaid predictions]{
    {\includegraphics[width=0.98\textwidth]{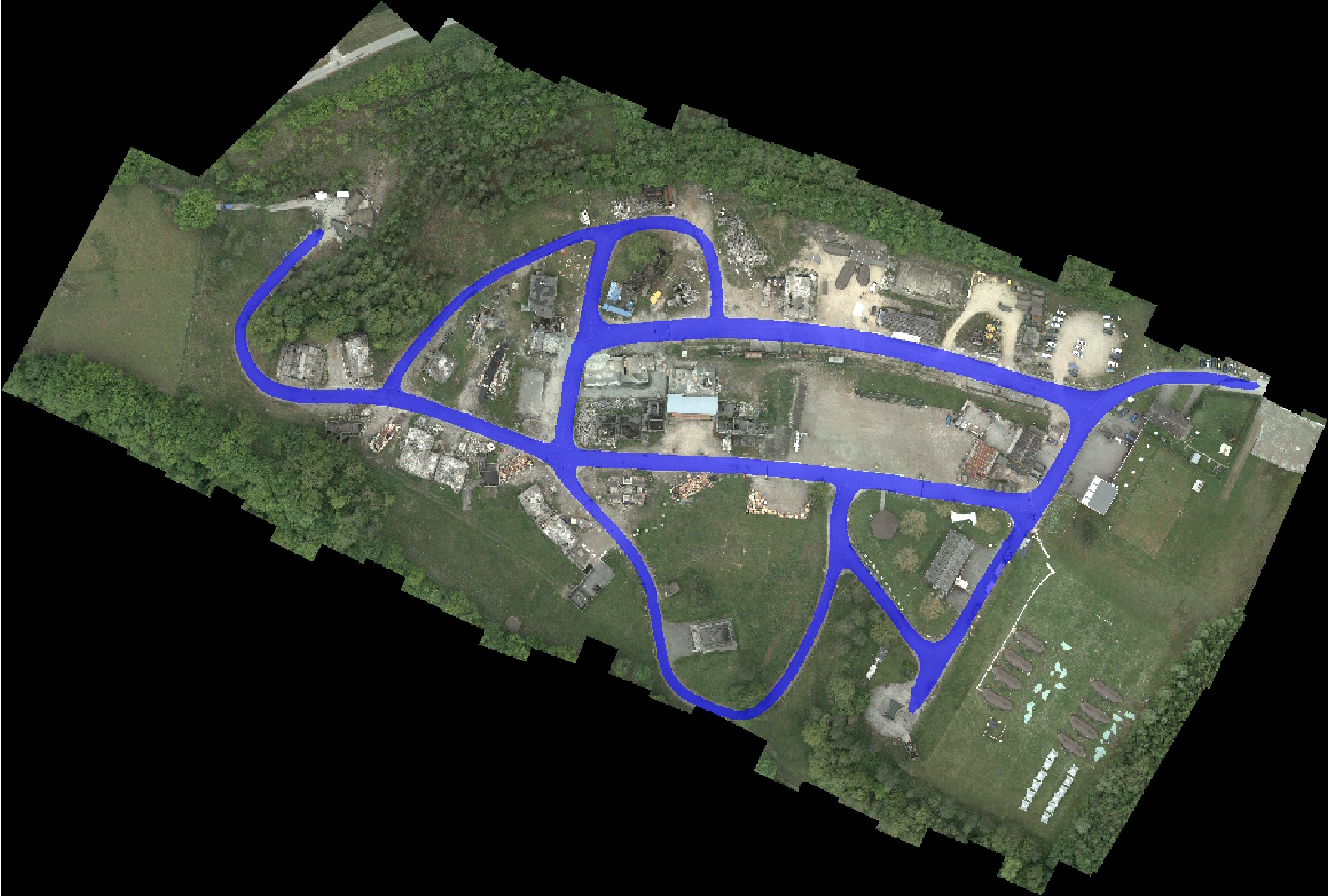}}}\\
    \vspace{-0.15cm}
    \subfigure[success case: zoom-in view of the image]{
    {\includegraphics[width=0.32\textwidth]{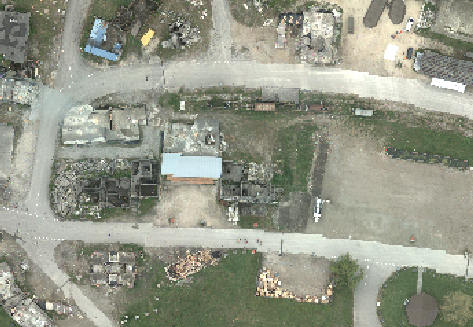}}}%
    \hspace{0.05cm}
    \subfigure[success case: zoom-in view of the predictions]{
    {\includegraphics[width=0.32\textwidth]{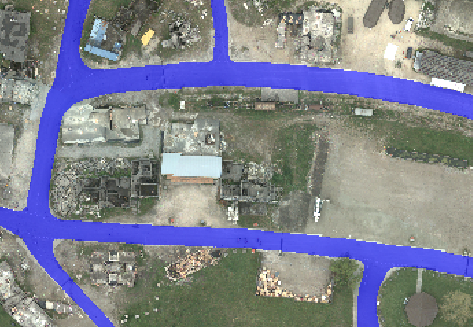}}}
    \hspace{0.05cm}
    \subfigure[success case: zoom-in view of the ground truth]{
    {\includegraphics[width=0.32\textwidth]{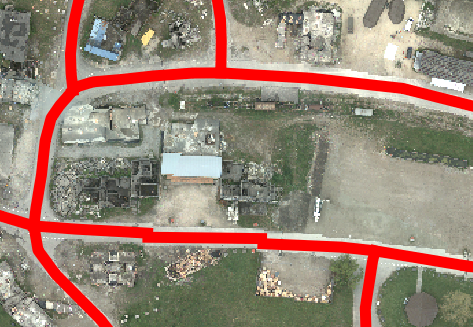}}}\\
    \vspace{-0.15cm}
    \subfigure[failure case: zoom-in view of the image]{
    {\includegraphics[width=0.32\textwidth]{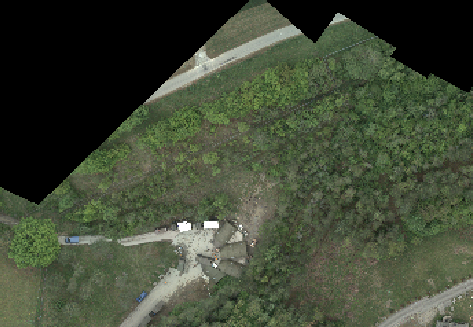}}}%
    \hspace{0.05cm}
    \subfigure[failure case: zoom-in view of the predictions]{
    {\includegraphics[width=0.32\textwidth]{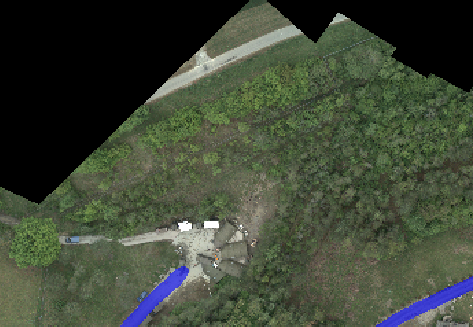}}}
    \hspace{0.05cm}
    \subfigure[failure case: zoom-in view of the ground truth]{
    {\includegraphics[width=0.32\textwidth]{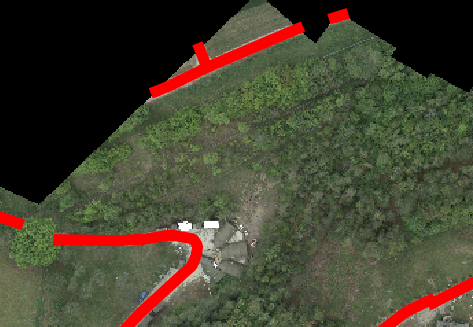}}}\\
    \vspace{-0.15cm}
    \caption{Road segmentation results for the test area in Epeisses, Switzerland.}
\label{Fig:appendix_road_epeisses}
\end{figure*}

\begin{figure*}[h!]
\centering 
    \subfigure[DOP20 pre-disaster image]{
        \includegraphics[width=0.48\textwidth]{
            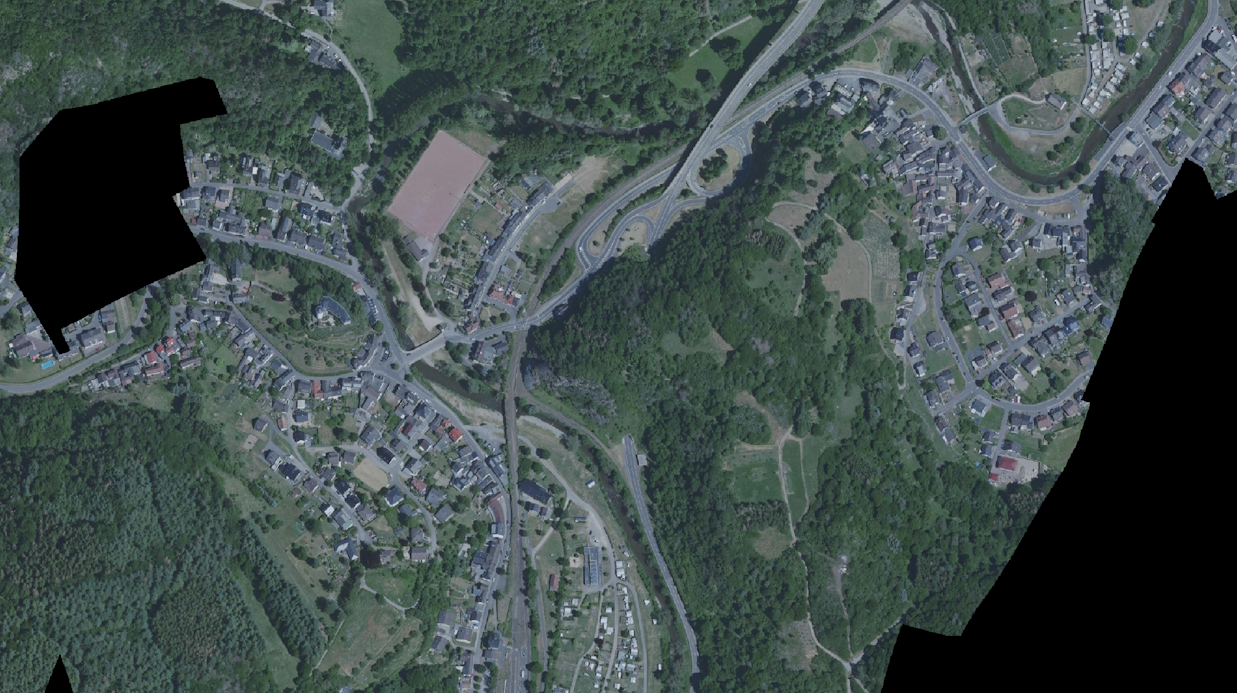
        }
    }
    \subfigure[4K post-disaster image]{
        \includegraphics[width=0.48\textwidth]{
            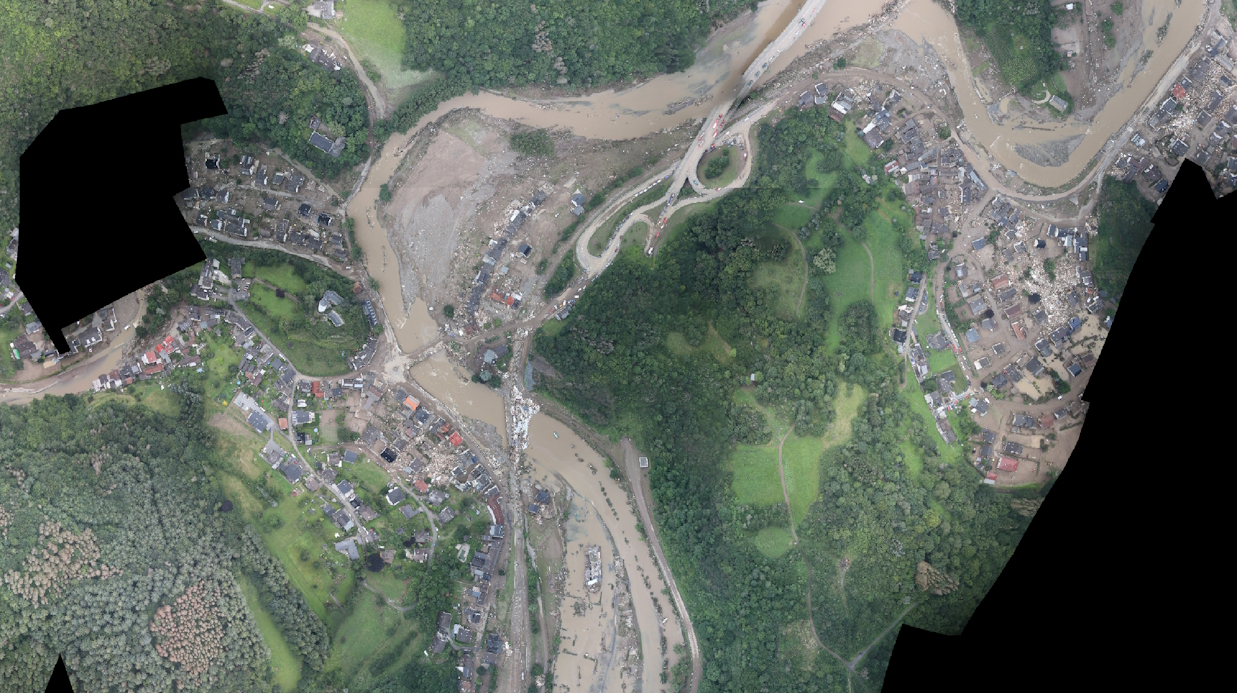
        }
    }
    \\
    \vspace{-0.15cm}
    \subfigure[DOP20 pre-disaster image with overlaid predictions]{
        \includegraphics[width=0.48\textwidth]{
            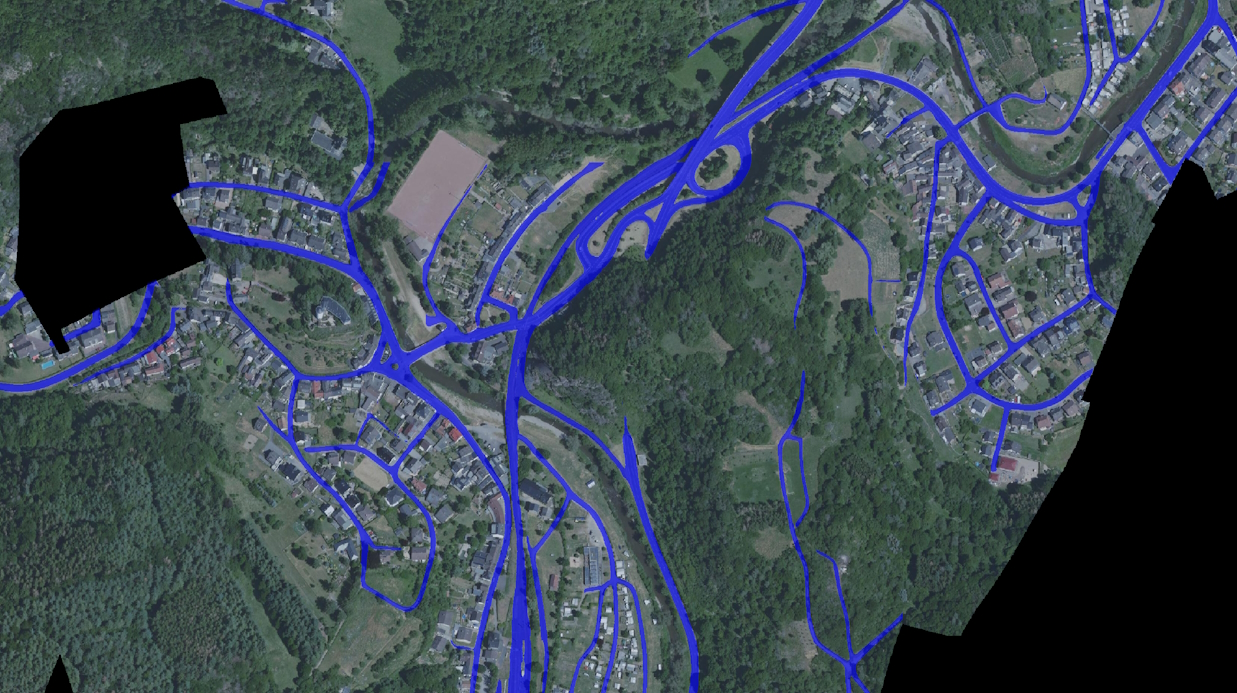
        }
    }
    \subfigure[4K post-disaster image with overlaid predictions]{
        \includegraphics[width=0.48\textwidth]{
            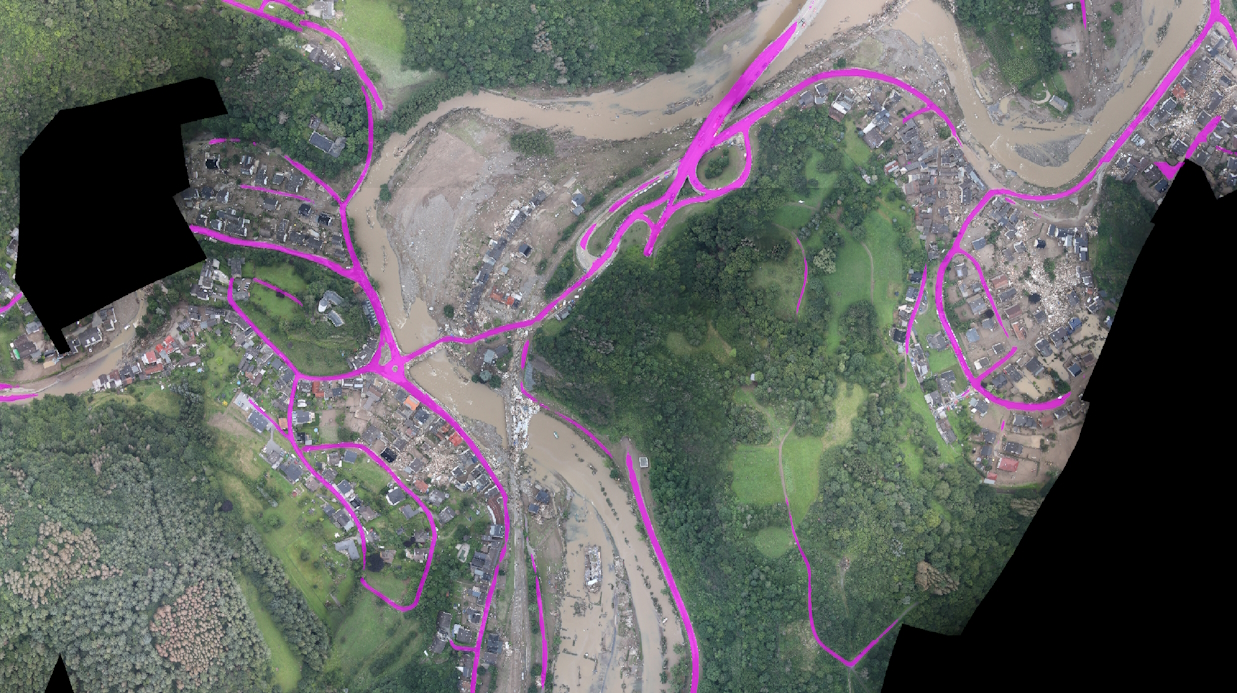
        }
    }
    \\
    \vspace{-0.15cm}
    \subfigure[zoom-in view of the DOP20 pre-disaster predictions]{
        \includegraphics[width=0.48\textwidth]{
            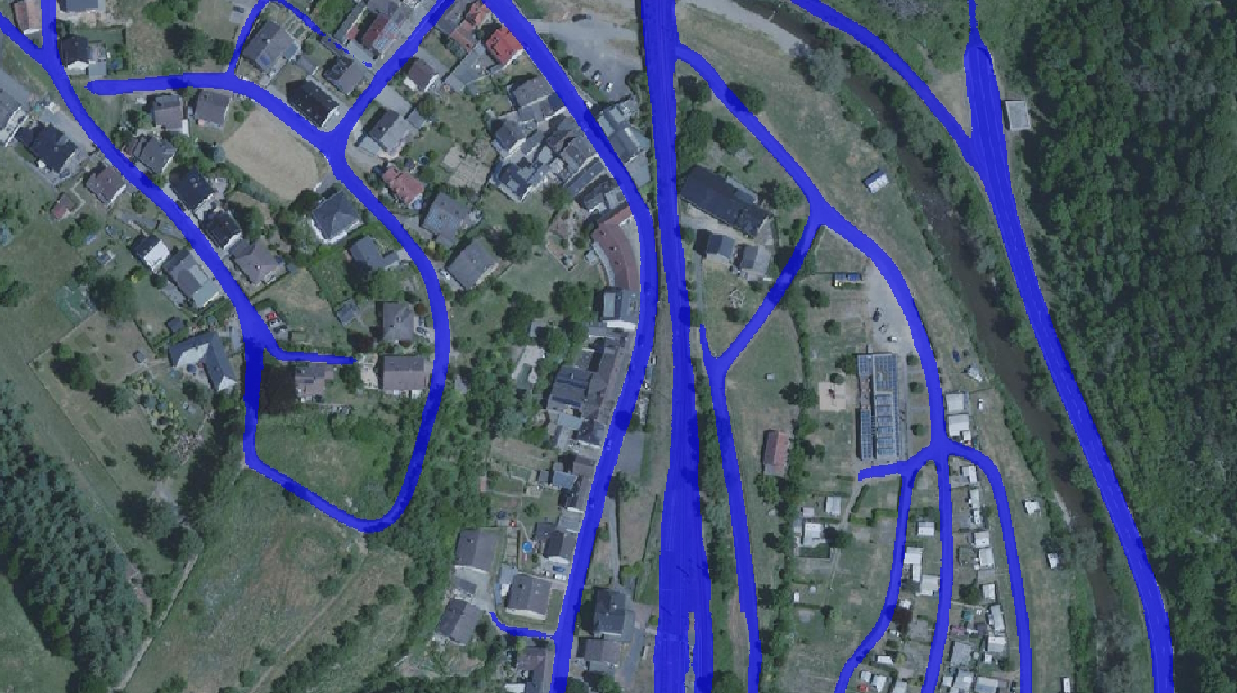
        }
    }
    \subfigure[zoom-in view of the 4K post-disaster predictions]{
        \includegraphics[width=0.48\textwidth]{
            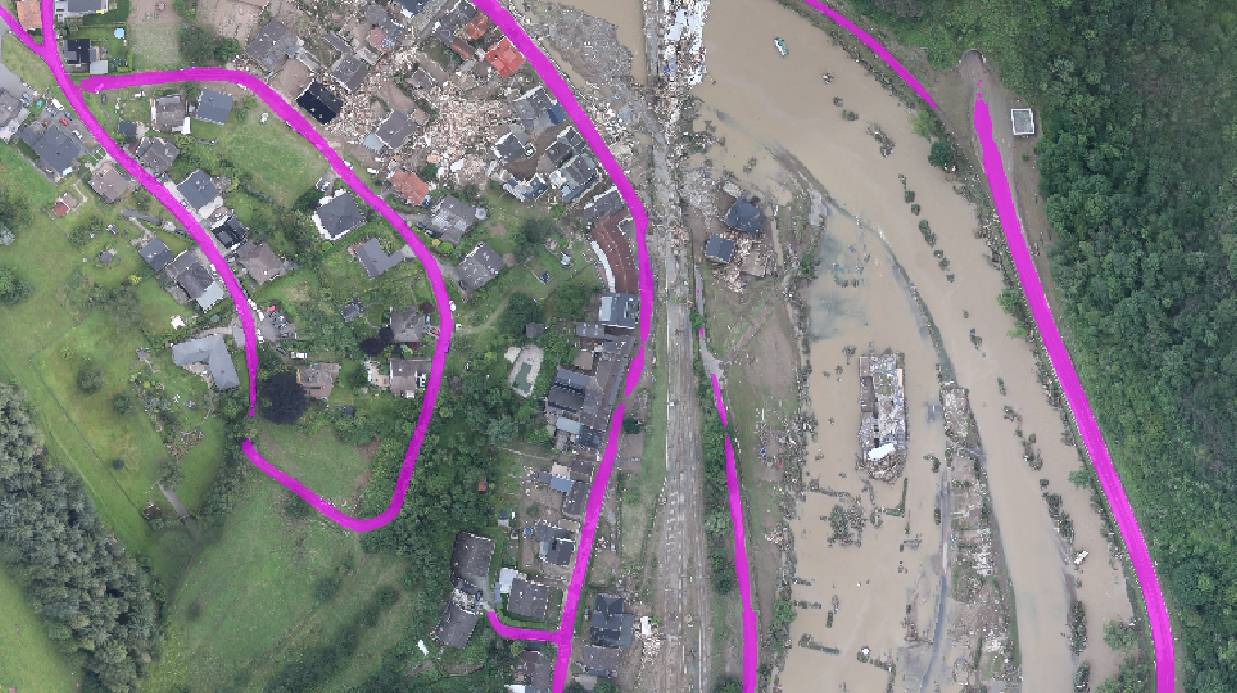
        }
    }
    \\
    \vspace{-0.15cm}
    \subfigure[zoom-in view of the DOP20 pre-disaster predictions]{
        \includegraphics[width=0.48\textwidth]{
            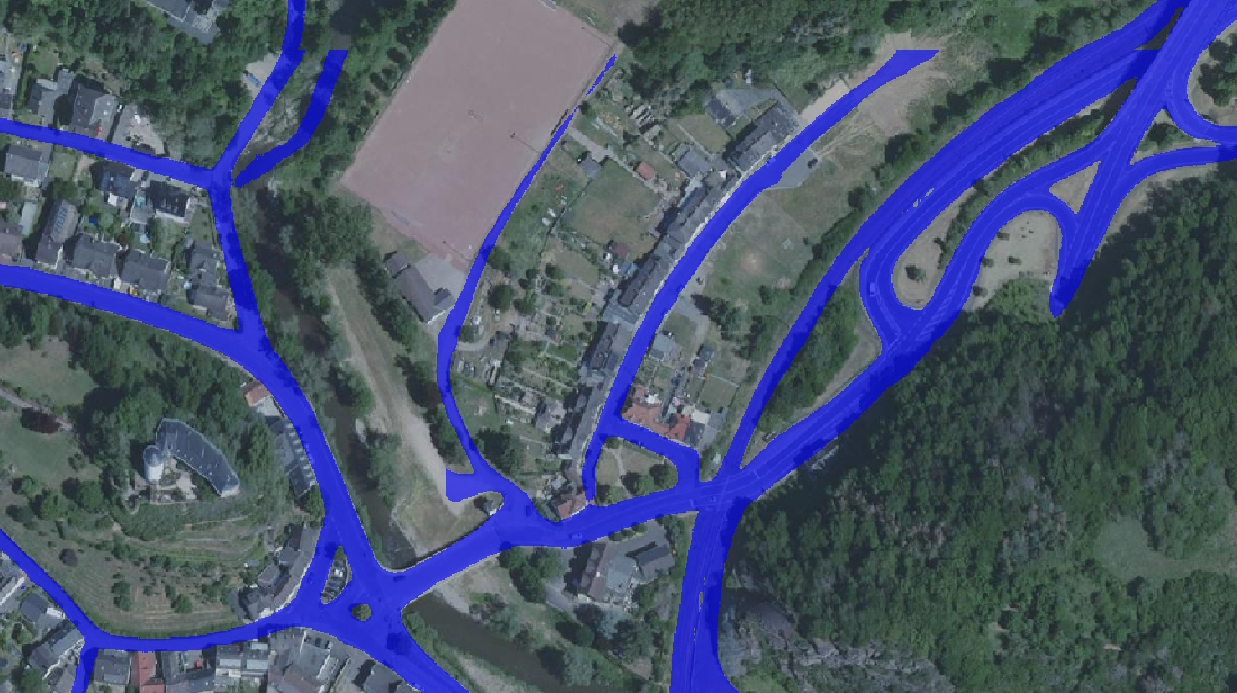
        }
    }
    \subfigure[zoom-in view of the 4K post-disaster predictions]{
        \includegraphics[width=0.48\textwidth]{
            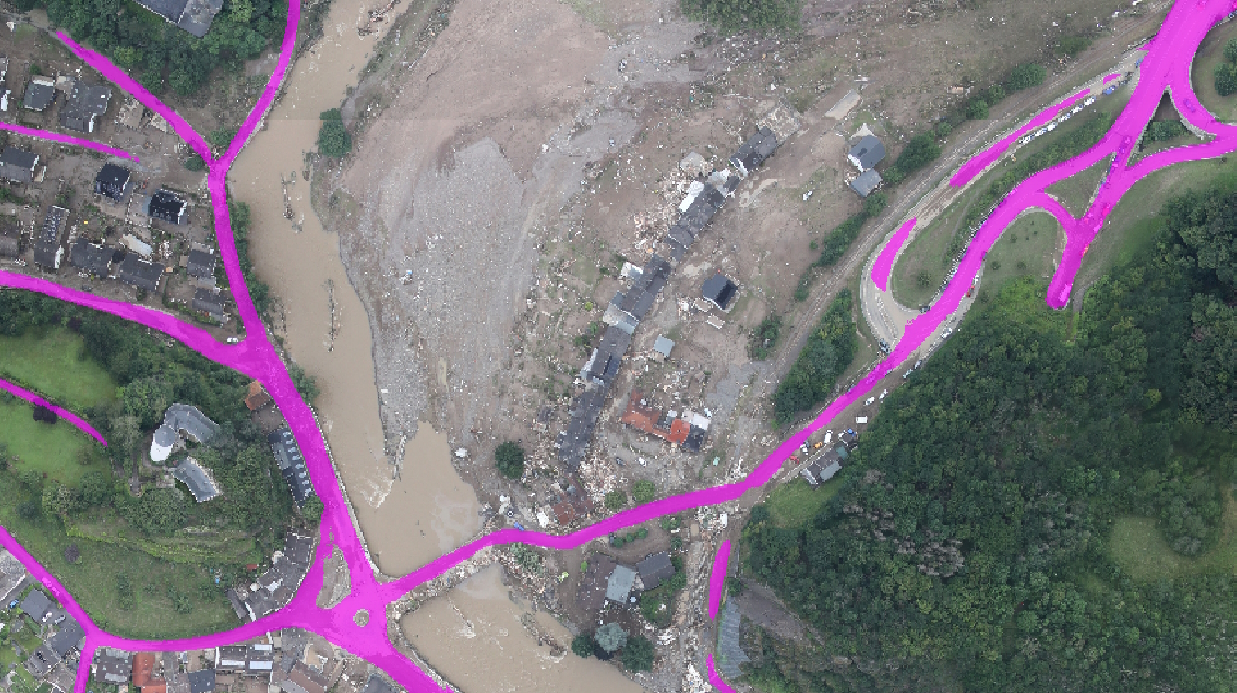
        }
    }
    \\
    \vspace{-0.15cm}
    \caption{Road segmentation results for the test areas in the Ahr Valley, Germany.}
\label{Fig:appendix_road_ahrtal}
\end{figure*}

\begin{figure*}[h!]
\centering 
    \subfigure[A selected scene from Beira with overlaid predictions]{
    {\includegraphics[width=0.75\textwidth]{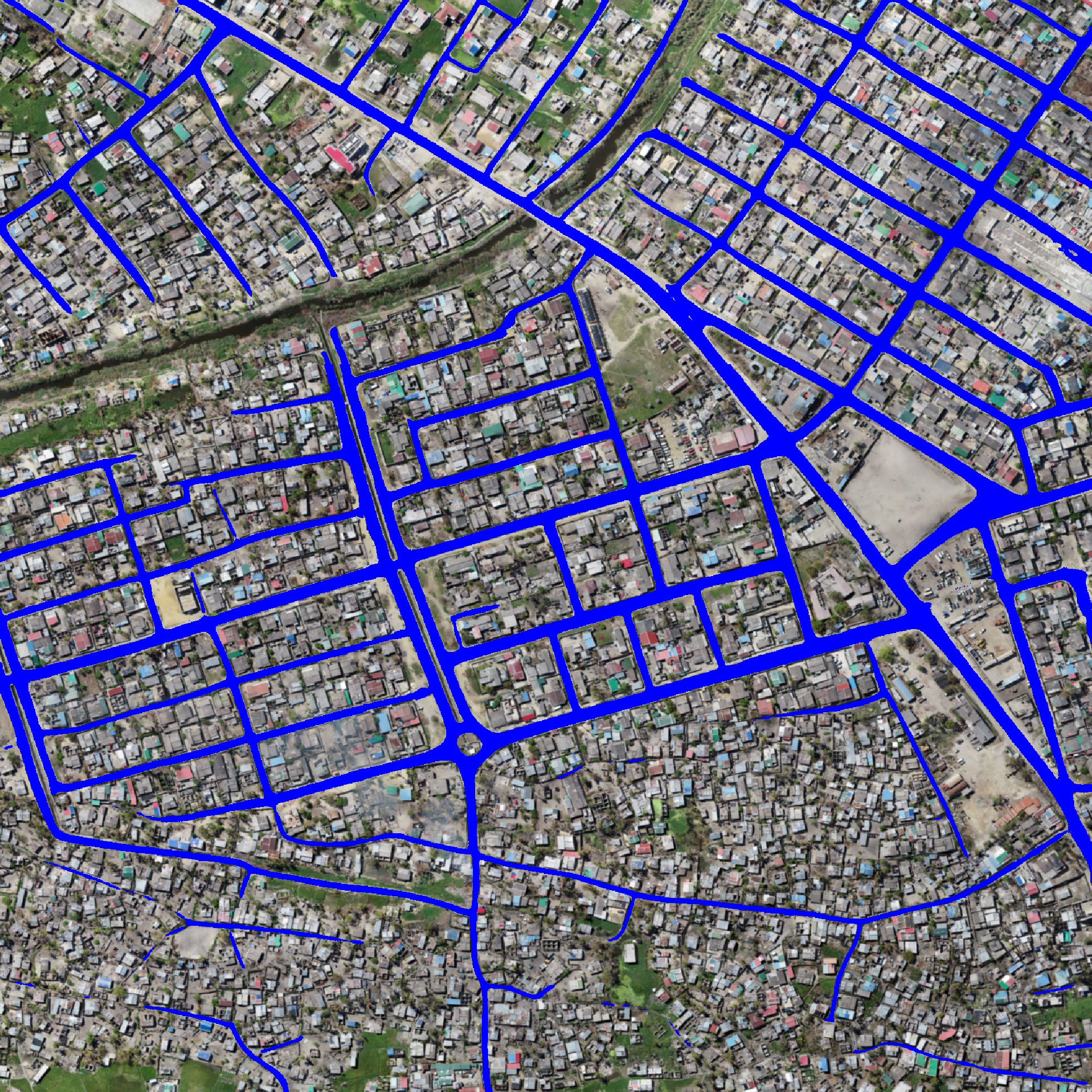}}}\\
    \vspace{-0.15cm}
    \subfigure[success case: zoom-in view of the image]{
    {\includegraphics[width=0.32\textwidth]{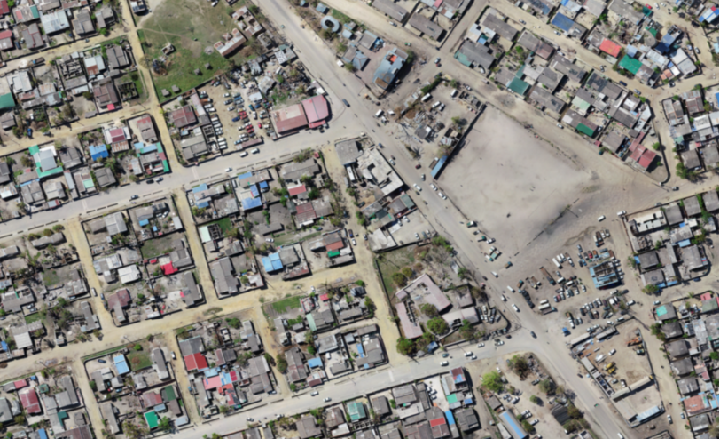}}}%
    \hspace{0.05cm}
    \subfigure[success case: zoom-in view of the predictions]{
    {\includegraphics[width=0.32\textwidth]{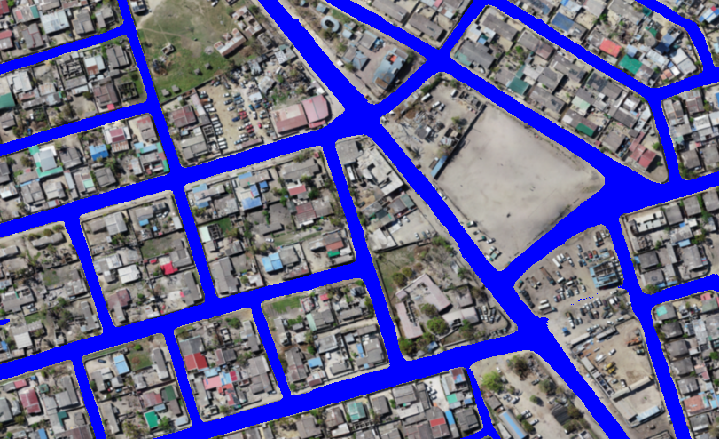}}}
    \hspace{0.05cm}
    \subfigure[success case: zoom-in view of the ground truth]{
    {\includegraphics[width=0.32\textwidth]{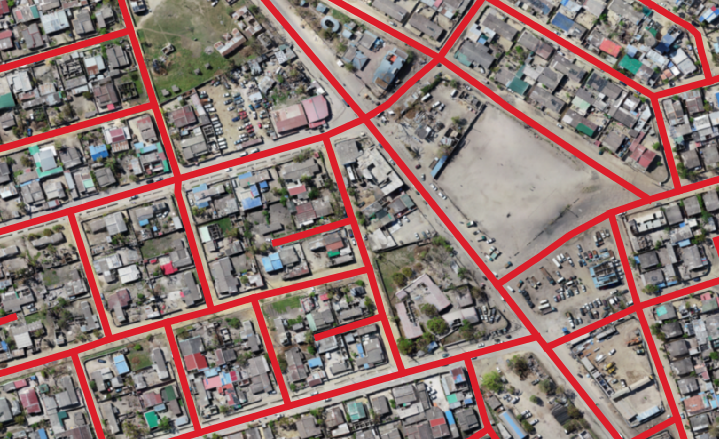}}}\\
    \vspace{-0.15cm}
    \subfigure[failure case: zoom-in view of the image]{
    {\includegraphics[width=0.32\textwidth]{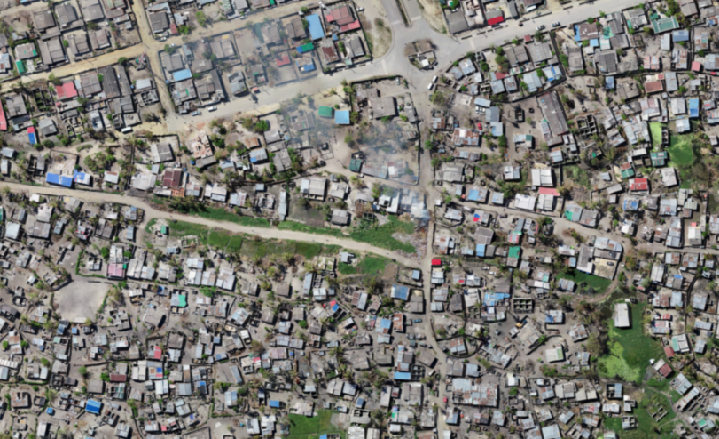}}}%
    \hspace{0.05cm}
    \subfigure[failure case: zoom-in view of the predictions]{
    {\includegraphics[width=0.32\textwidth]{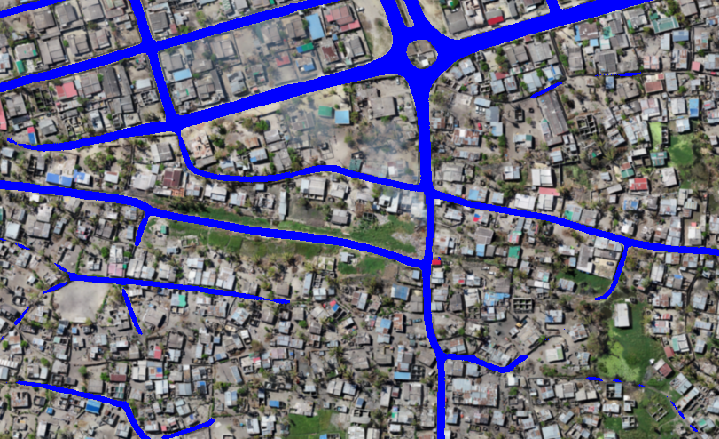}}}
    \hspace{0.05cm}
    \subfigure[failure case: zoom-in view of the ground truth]{
    {\includegraphics[width=0.32\textwidth]{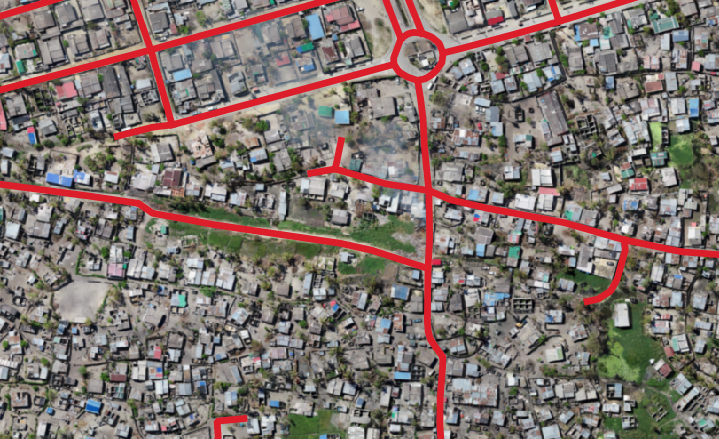}}}\\
    \vspace{-0.15cm}
    \caption{Road segmentation results for a scene of the test areas in Beira, Mozambique.}
\label{Fig:appendix_road_beira}
\end{figure*}

\begin{figure*}[h!]
\centering 
    \includegraphics[width=\textwidth]{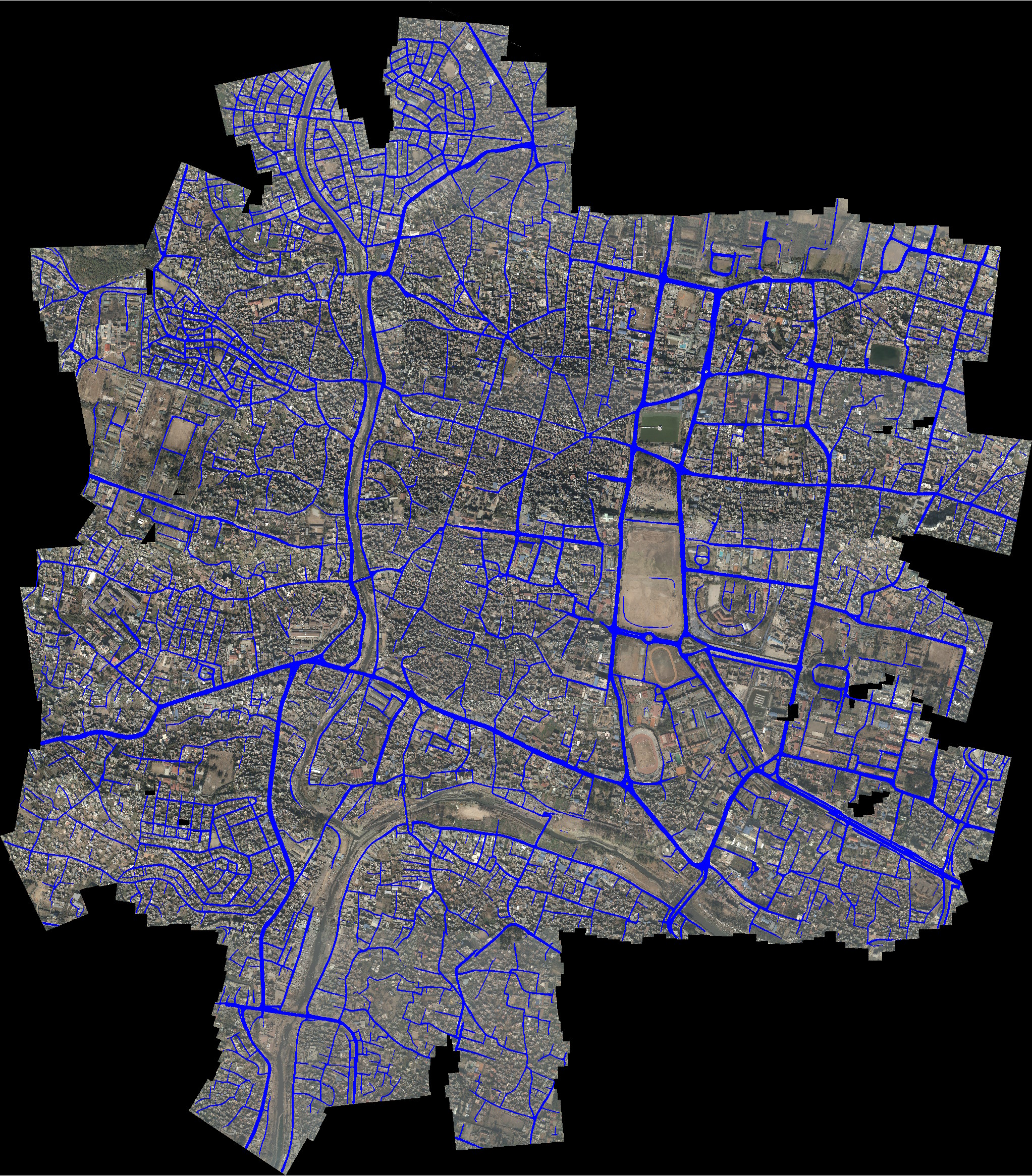}\\
    \vspace{-0.15cm}
    \caption{Road segmentation results for a scene from Kathmandu in Nepal captured from a UAV at 8~cm/px and resampled to 50~cm/px.}
\label{Fig:appendix_road_kathmandu}
\end{figure*}

We tested the road segmentation method on 21 test areas from Epeisses in Switzerland, from the Ahr Vally in Germany, from Beira in Mozambique, and from Kathmandu in Nepal. For all scenes except Kathmandu, we annotated the images by hand following the centerline of the roads and saved them as vector graphics. \autoref{Fig:appendix_road_epeisses}, \autoref{Fig:appendix_road_ahrtal}, \autoref{Fig:appendix_road_beira} and \autoref{Fig:appendix_road_kathmandu} show the predictions of our model and a dilated version of our ground truth if available. Note that although our model outputs pixel-wise segmentation, our metrics do not compare them pixel to pixel to the labels as shown in the figures. Rather, they are evaluated on a topological basis after the predictions have been thinned to a 1-pixel thickness, equivalent to vectorizing them into centerlines.

\textbf{In the Epeisses scene} (see \autoref{Fig:appendix_road_epeisses}), most roads were correctly identified and accurately extracted, except for some sections located close to the edges of the mosaic. This is due to the lack of context given to the model, which expects the roads to be continuous as it has not been trained to overcome sudden disruptions by the background areas in the images.

\textbf{In the Ahr Valley scene} (see \autoref{Fig:appendix_road_ahrtal}), the model managed to detect most roads in both the pre- and post-disaster images, though we could not report results as no ground truth was available for this area yet. In the pre-disaster image, our model shows its capacity for generalizing well to sub-urban scene types unseen during its training, as it was only given to see regions from Southeast Asia. In the post-disaster image, it has shown some confusion as to which road to consider as still intact: there is in fact much water, mud, and debris on the surface of the roads, making it more difficult to draw a line between damaged and usable road sections.

\textbf{In the Beira scene} (see \autoref{Fig:appendix_road_beira}), while the images are particularly challenging due to the presence of unpaved roads or streets covered in sand in the aftermath of Cyclone Idai, the model still manages to extract all the roads except a few narrower ones. However, it did detect roads that the annotators did not include in the labels due to occlusion or the lack of clues as to their usability by vehicles. This begs the question of the annotation policy and the boundaries between road and non-road objects in difficult scenarios where they might either be not visible or require local knowledge for a specific region.

\textbf{In the Kathmandu scene} (see \autoref{Fig:appendix_road_kathmandu}), the model was faced with a complex urban infrastructure, featuring many narrow, irregular, and therefore occluded streets, and often unpaved roads. Nevertheless, it was capable of extracting the vast majority of the roads with great accuracy, from large arteries to small alleyways, even though the connectivity of the mask may be improved in locations where the road segments are kept apart by the occlusion of buildings. In such scenarios, it actually becomes challenging to define a fair, comprehensive road annotation policy, as expert on-the-ground knowledge is required to define the difference between a road and a simple large- drivable area not dedicated to vehicles.

\subsection{Detailed results for building segmentation}

The building segmentation method is tested on 15 areas, including those mentioned in the main text plus one area in Kathmandu, Nepal (same as for the road segmentation).

\begin{figure*}[htbp!]
\centering 
    \subfigure[The complete scene in Epeisses with overlaid predictions]{
    {\includegraphics[width=0.98\textwidth]{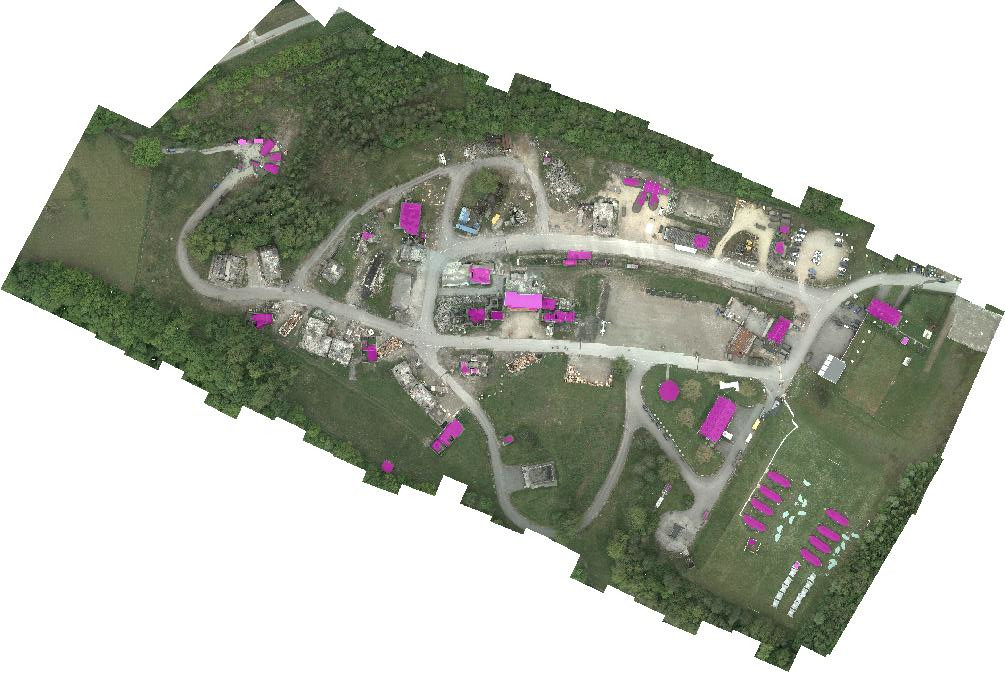}}}\\
    \vspace{-0.15cm}
    \subfigure[failure case: zoom-in view of the image]{
    {\includegraphics[width=0.32\textwidth]{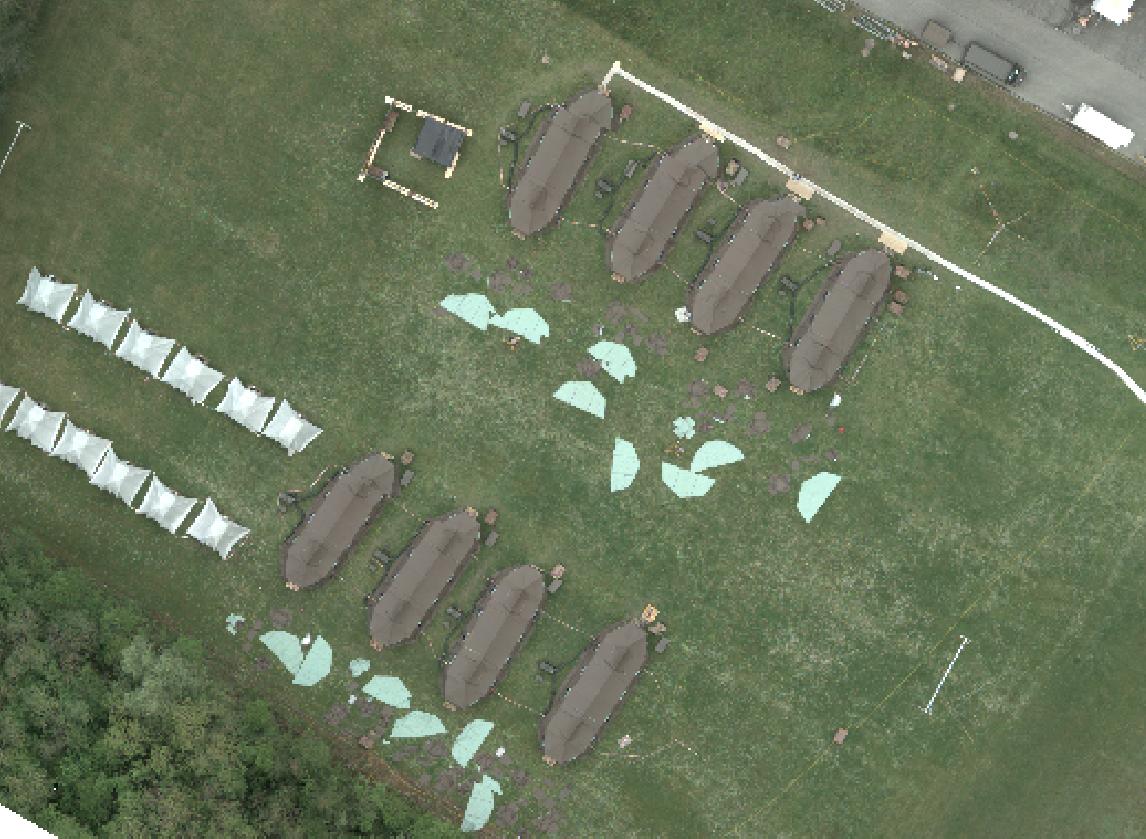}}}%
    \hspace{0.05cm}
    \subfigure[failure case: zoom-in view of the predictions]{
    {\includegraphics[width=0.32\textwidth]{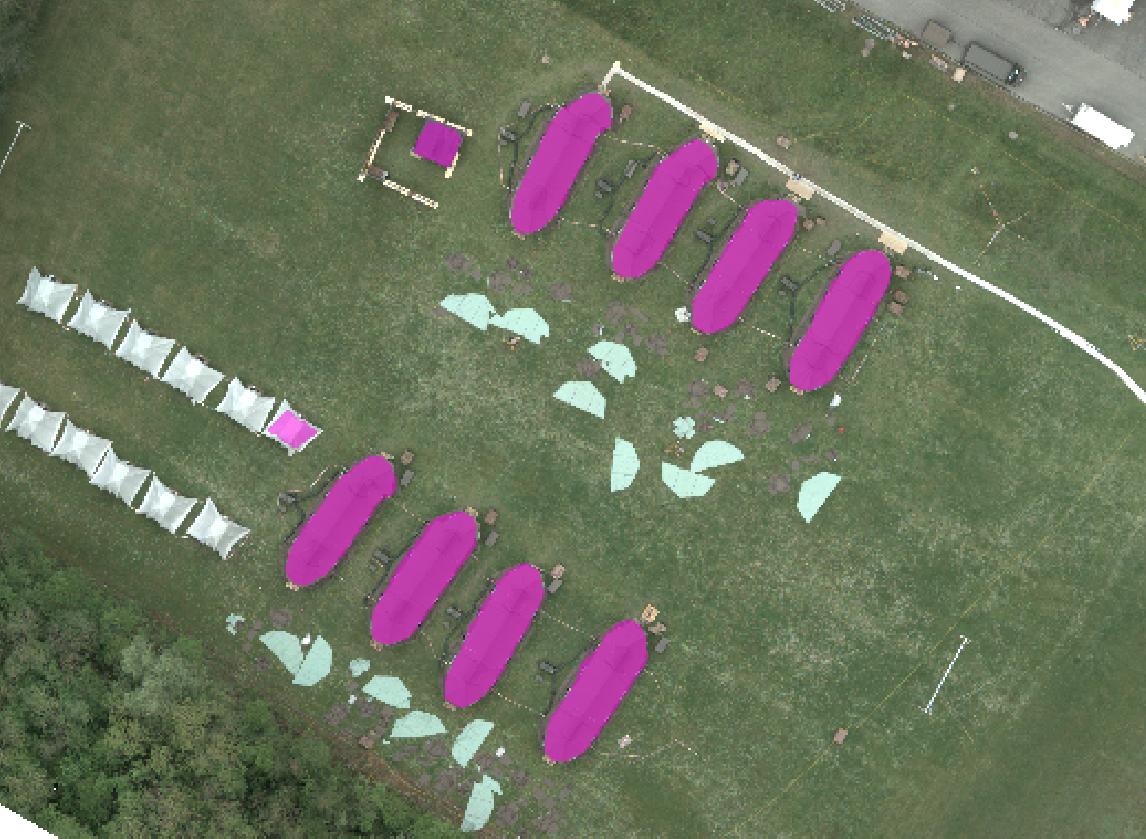}}}
    \hspace{0.05cm}
    \subfigure[failure case: zoom-in view of the ground truth]{
    {\includegraphics[width=0.32\textwidth]{Images/appendix/building/epeisses_zoomin1_img.jpg}}}\\
    \vspace{-0.15cm}
    \subfigure[failure case: zoom-in view of the image]{
    {\includegraphics[width=0.32\textwidth]{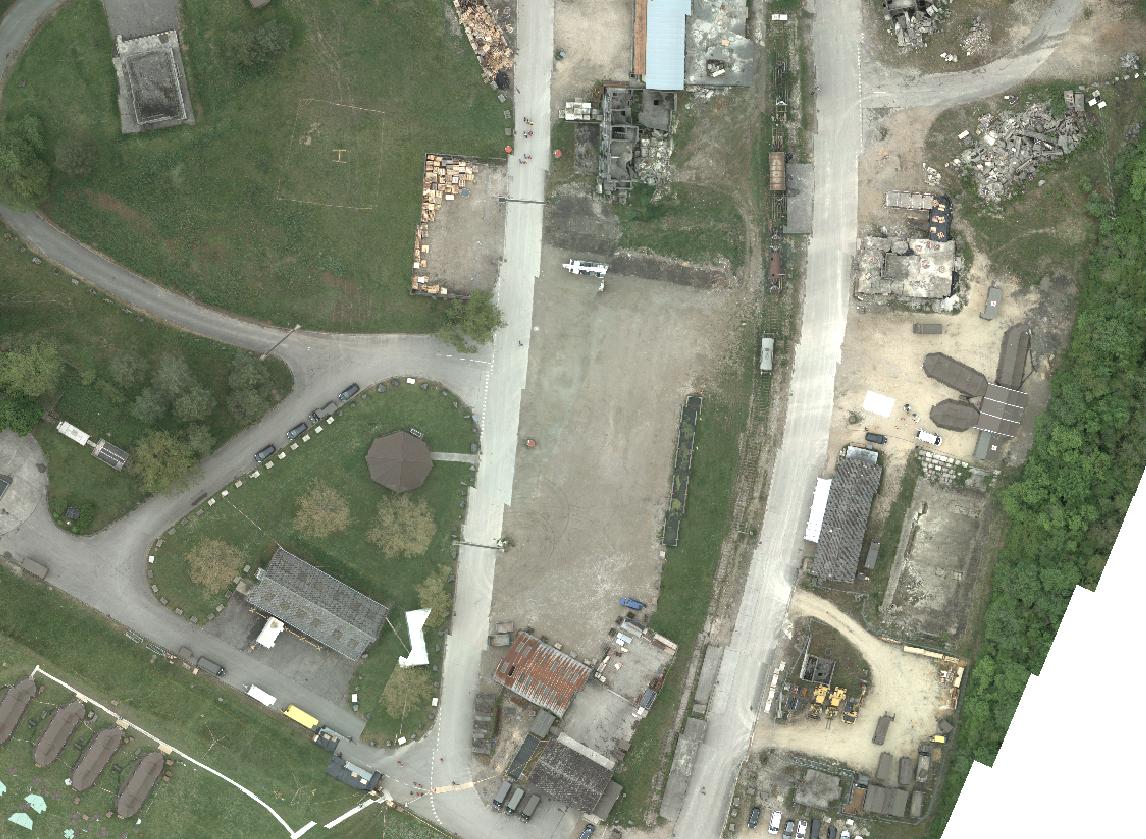}}}%
    \hspace{0.05cm}
    \subfigure[failure case: zoom-in view of the predictions]{
    {\includegraphics[width=0.32\textwidth]{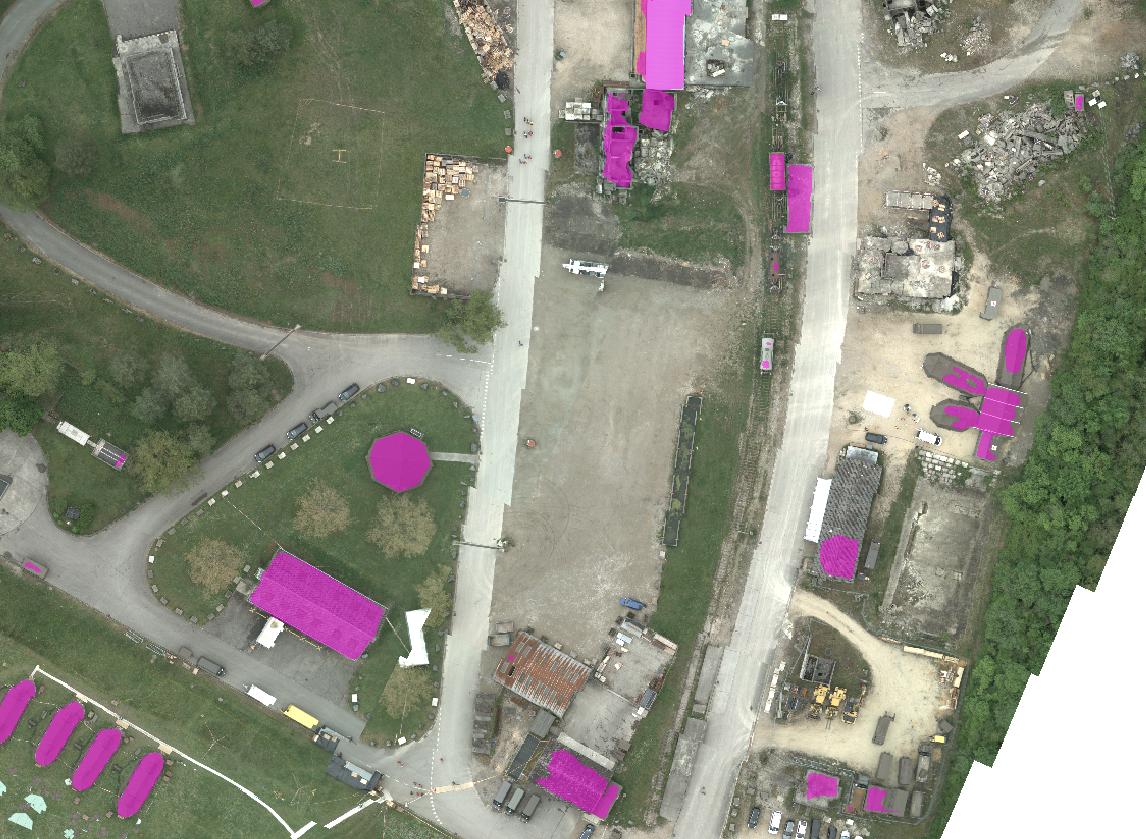}}}
    \hspace{0.05cm}
    \subfigure[failure case: zoom-in view of the ground truth]{
    {\includegraphics[width=0.32\textwidth]{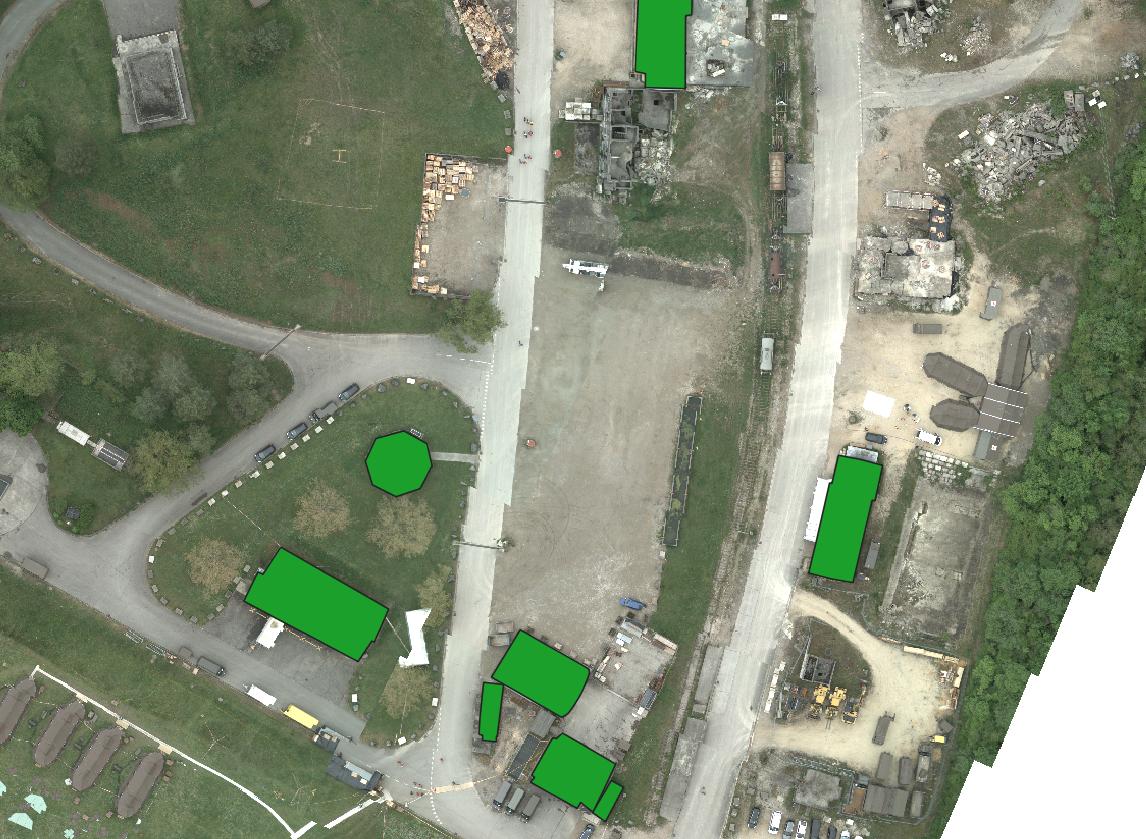}}}\\
    \vspace{-0.15cm}
    \caption{Building segmentation results for the test area in Epeisses, Switzerland.}
\label{Fig:appendix_building_epeisses}
\end{figure*}

\begin{figure*}[!ht]
\centering 
    \subfigure[DOP20 pre-disaster image with overlaid predictions]{
    {\includegraphics[width=0.48\textwidth]{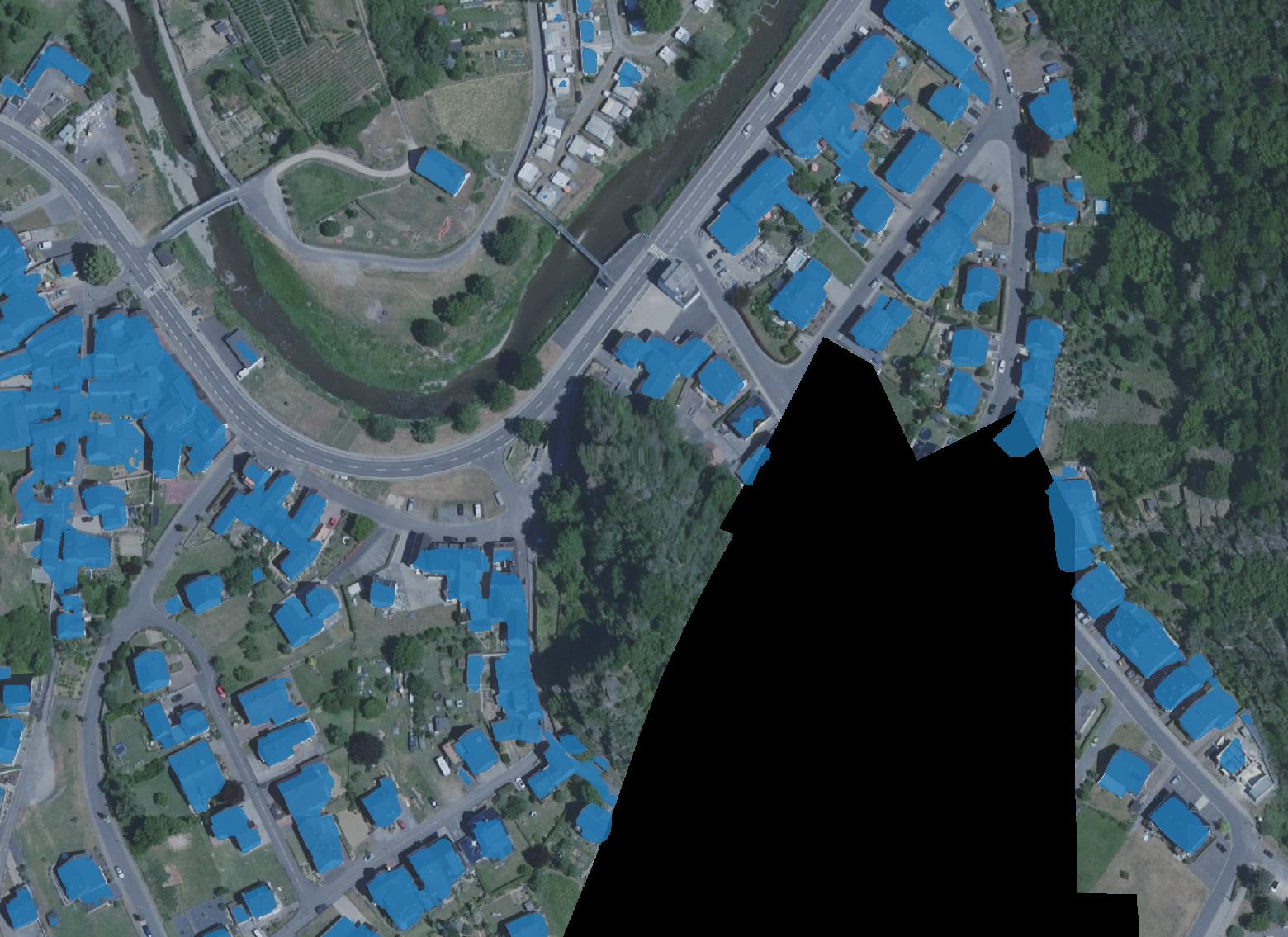}}}
    \subfigure[MACS post-disaster image with overlaid predictions]{
    {\includegraphics[width=0.48\textwidth]{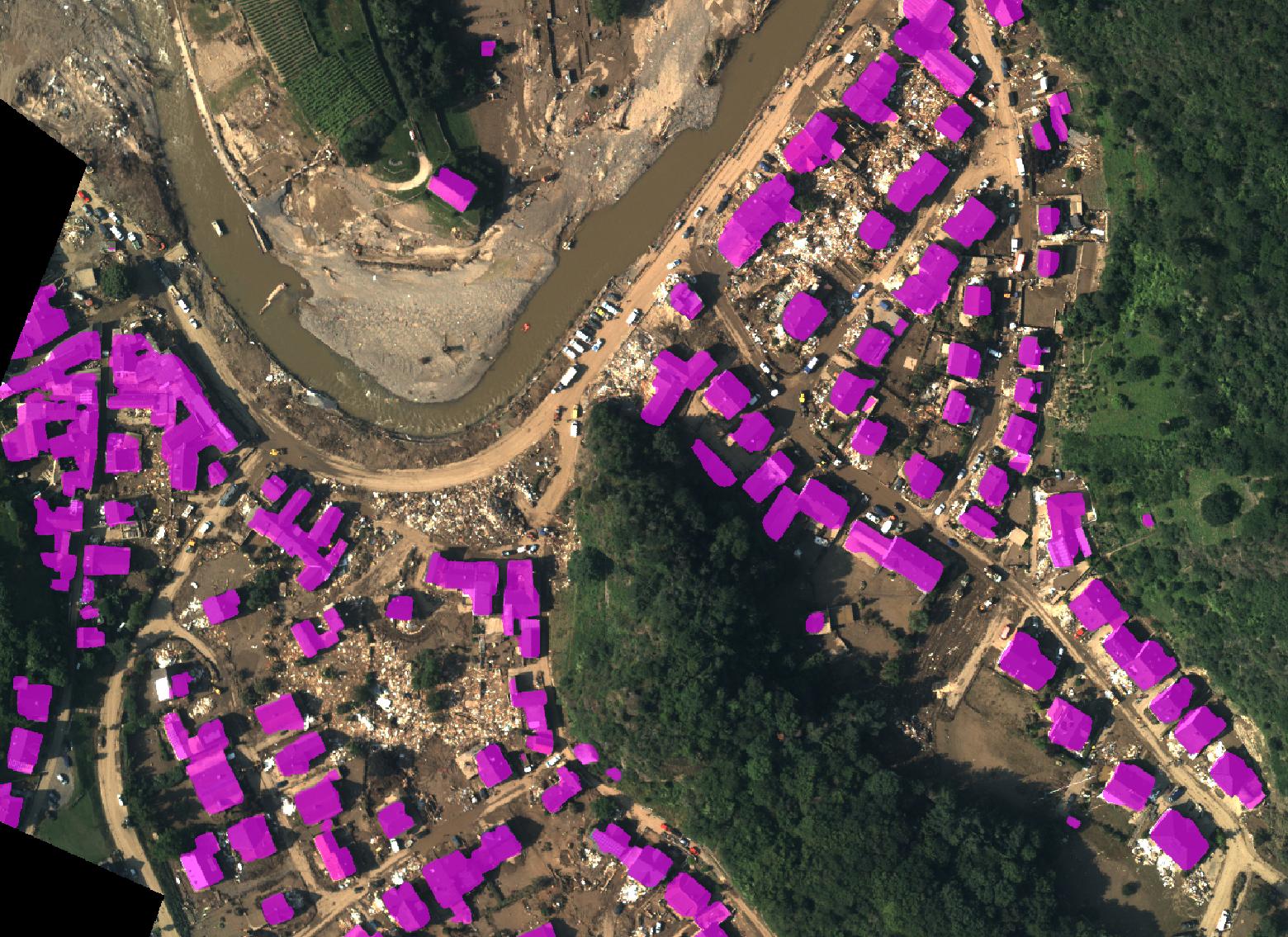}}}
    \\
    
    \vspace{-0.15cm}
    
    \subfigure[DOP20 pre-disaster image with overlaid predictions]{
    {\includegraphics[width=0.48\textwidth,trim={0 150 300 0},clip]{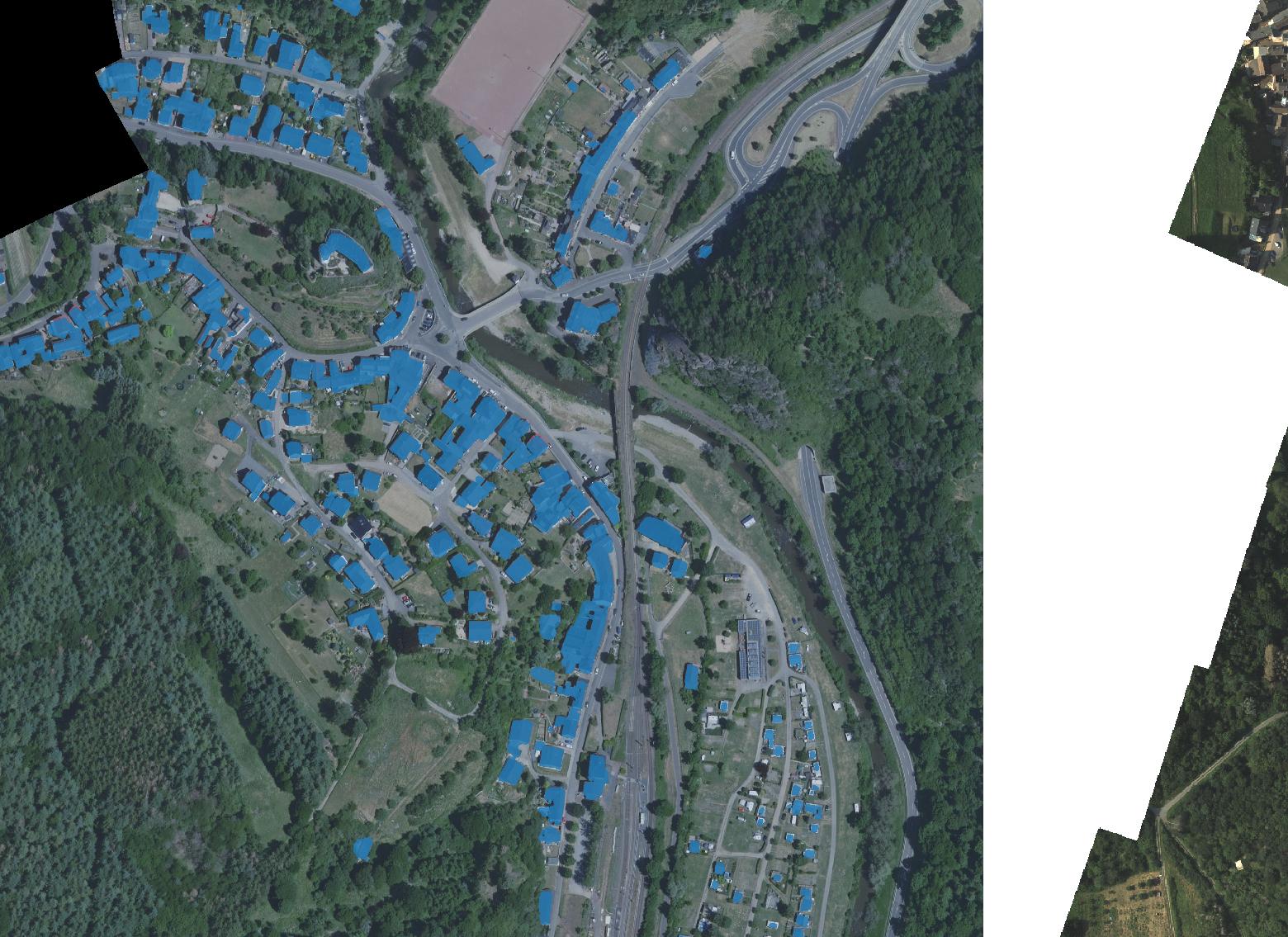}}}%
    \hspace{0.05cm}
    \subfigure[4K post-disaster image with overlaid predictions]{
    {\includegraphics[width=0.48\textwidth,trim={0 150 300 0},clip]{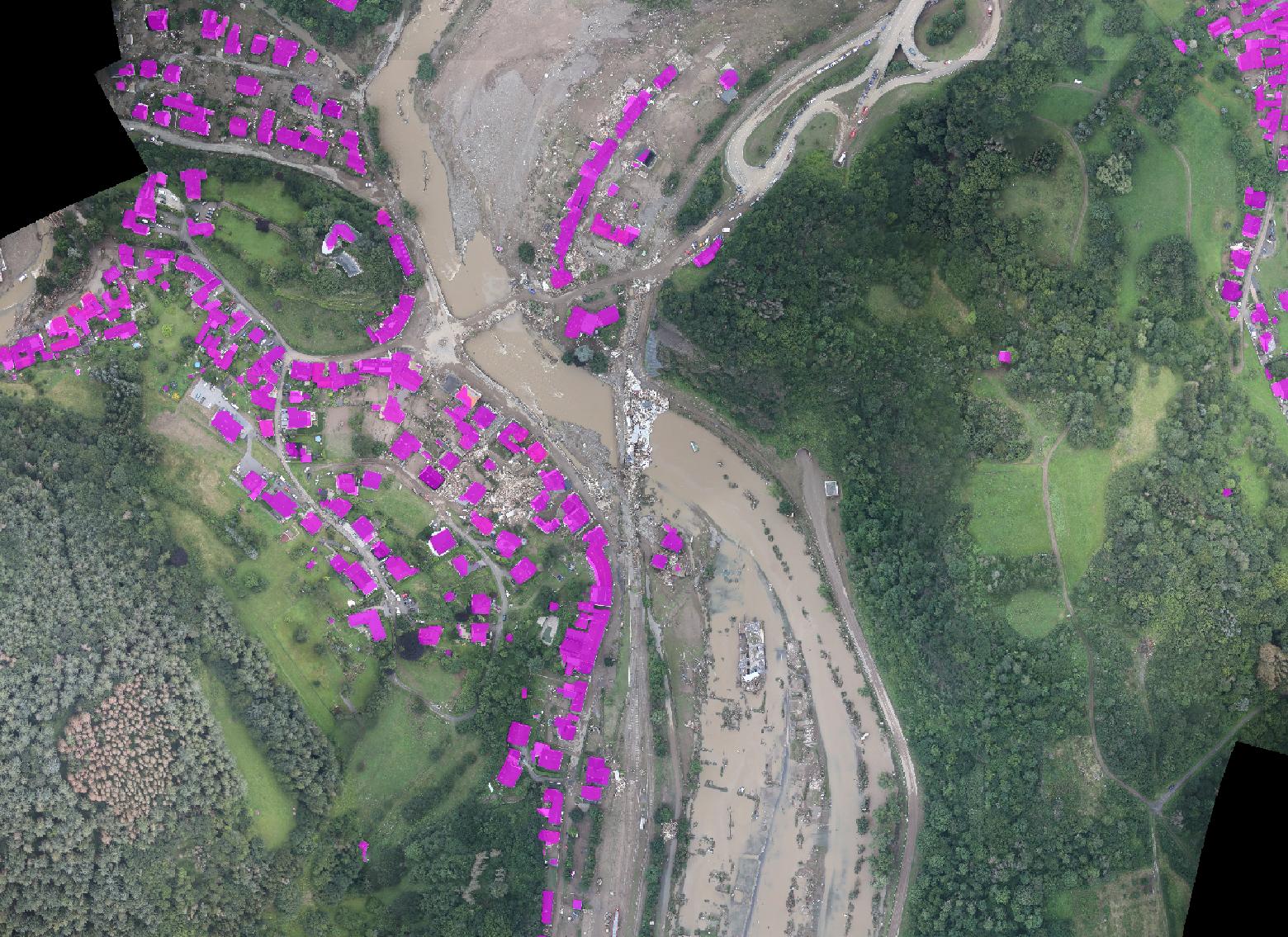}}}
    \\
    
    \vspace{-0.15cm}

    \subfigure[DOP20 pre-disaster image with overlaid predictions]{
    {\includegraphics[width=0.48\textwidth]{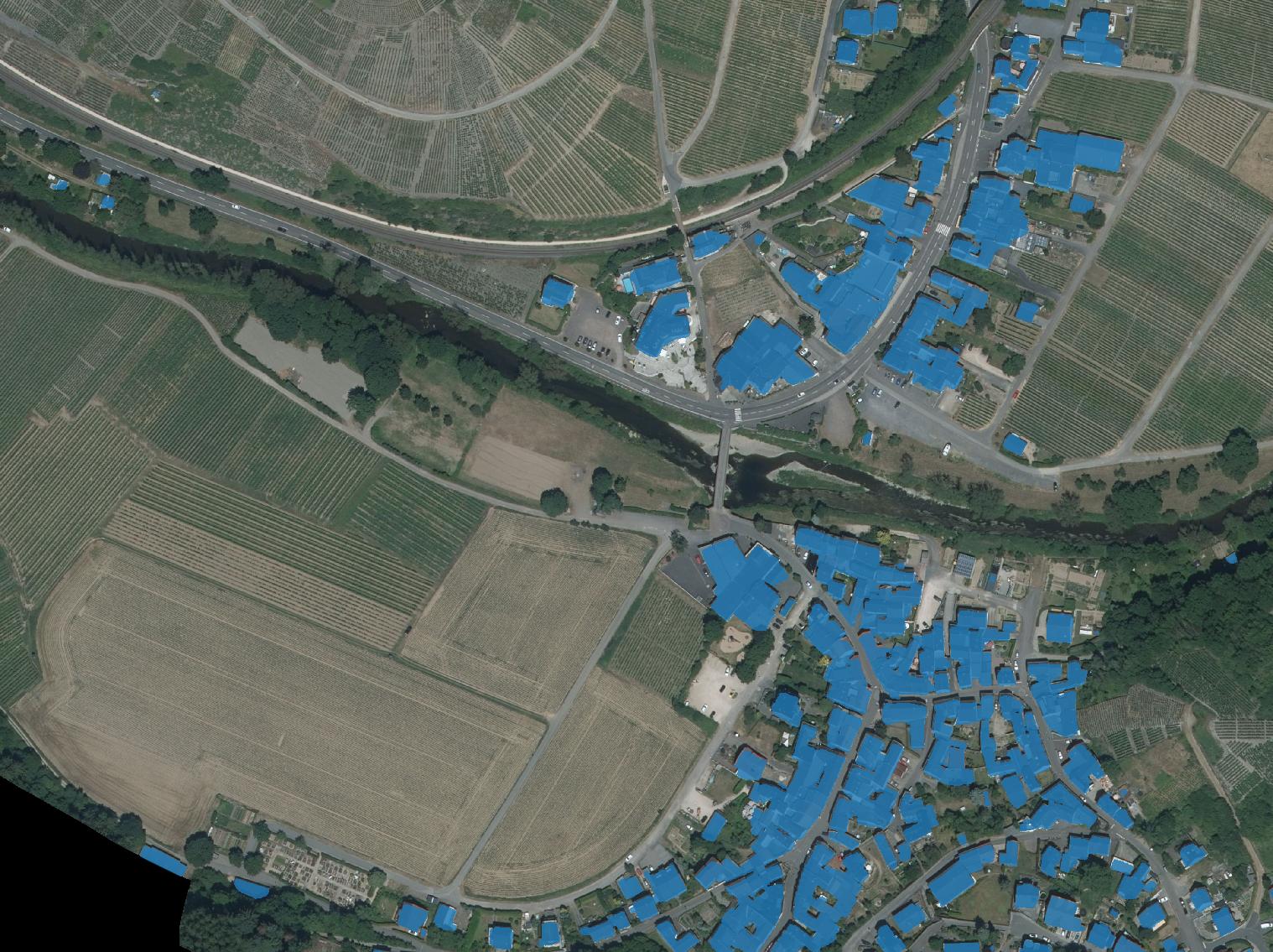}}}%
    \hspace{0.05cm}
    \subfigure[4K post-disaster image with overlaid predictions]{
    {\includegraphics[width=0.48\textwidth]{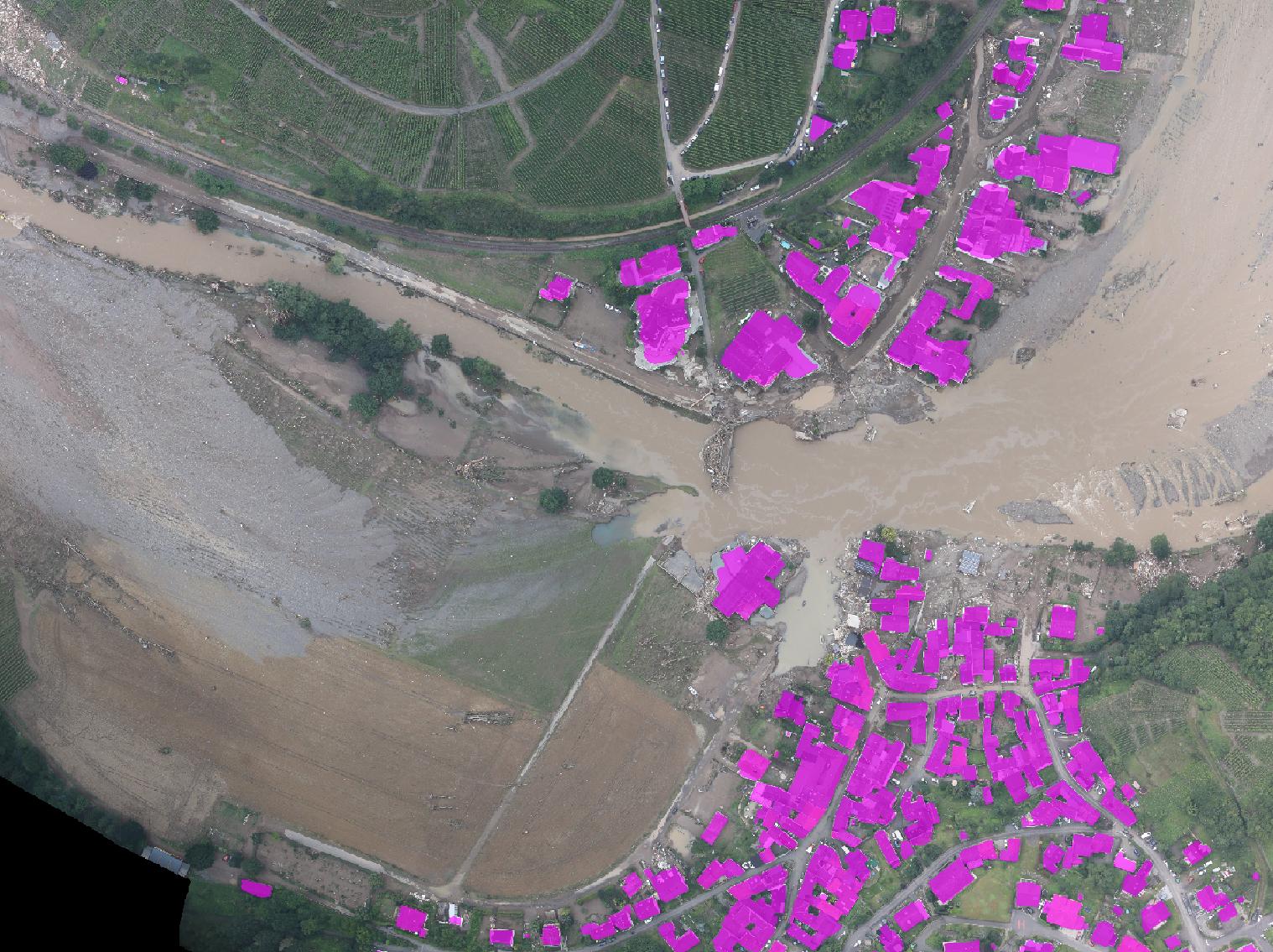}}}
    \\
    
    \vspace{-0.15cm}
    \caption{Building segmentation results of the selected areas in the Ahr Valley, Germany.}
\label{Fig:appendix_building_ahrtal}
\end{figure*}

\textbf{In the Epeisses scene} (see \autoref{Fig:appendix_building_epeisses}), most of the buildings are extracted and only two buildings are omitted. On the other side, large tents are mistakenly segmented due to their similarity to real buildings. In addition, damaged and collapsed buildings are classified as buildings but are not labeled in the annotation. Overall, an F1 and IoU score of 47.72\% and 31.33\% are acquired. 
\textbf{In the Ahrtal scene} (see \autoref{Fig:appendix_building_ahrtal}), we selected 10 regions (10.5km$^2$) and manually annotated the building ground truth of the pre-flood images. In the 10 annotated regions we achieved 86.66\% and 76.46\% for building F1 and IoU scores respectively. \autoref{Fig:appendix_building_ahrtal} illustrates three small regions with pre- and post-flood images. All pre-event images are DOP20, the post-event image (b) is captured by the MACS camera system, and (d) and (f) illustrate images captured by the 4k system. Due to the difference in flight altitude and viewing angles between the training and test data, the network was unable to detect a couple of damaged buildings from a very oblique view (see \autoref{Fig:appendix_building_ahrtal}~(d)), but still managed to accurately extract most of the buildings.

\begin{figure*}[htpb!]
\centering 
    \subfigure[A larger scene selected from Beira, Mozambique with overlaid predictions]{
    {\includegraphics[width=0.96\textwidth]{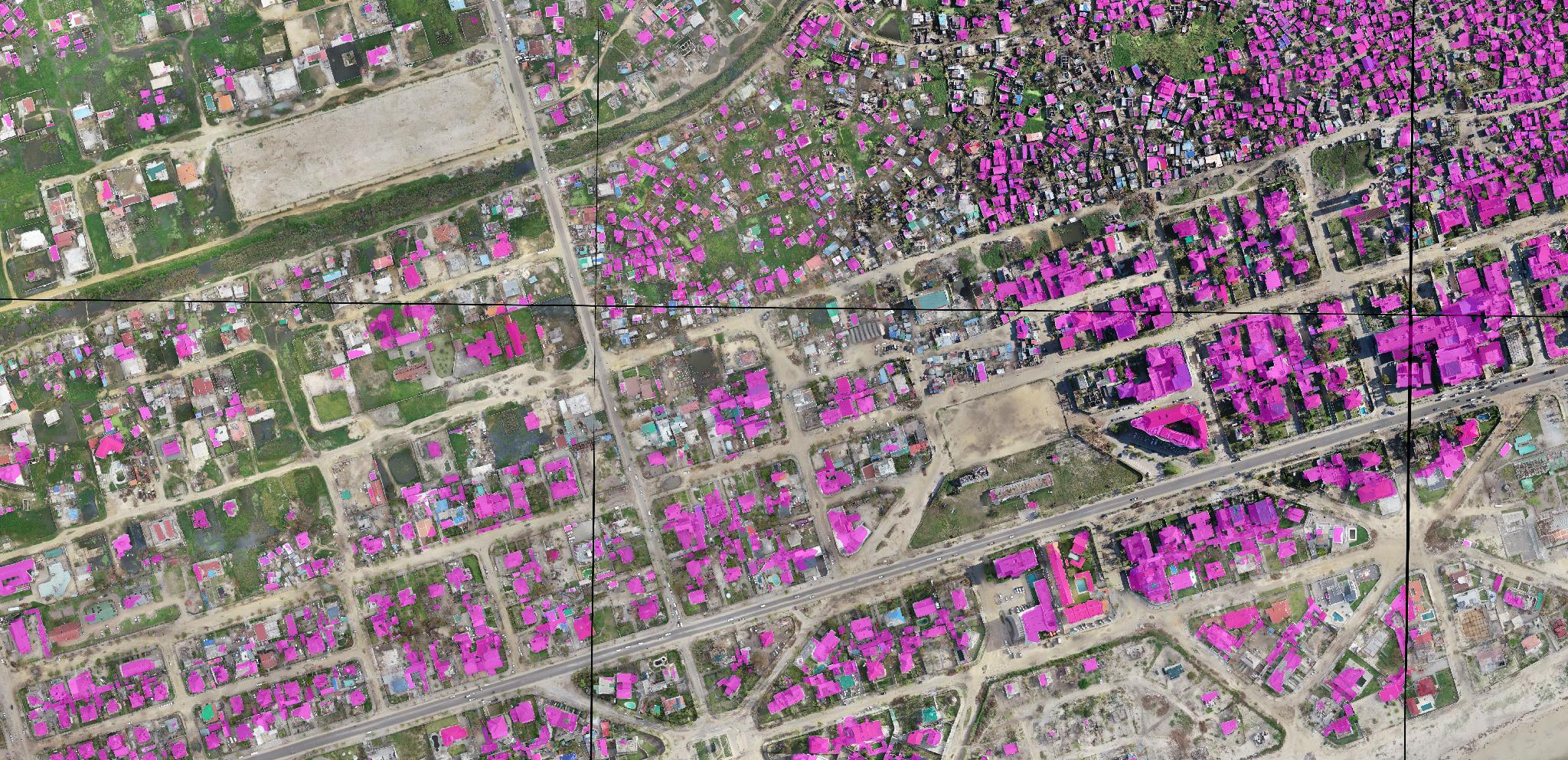}}}\\
    \vspace{-0.15cm}
    \subfigure[zoom-in view of the image]{
    {\includegraphics[width=0.32\textwidth]{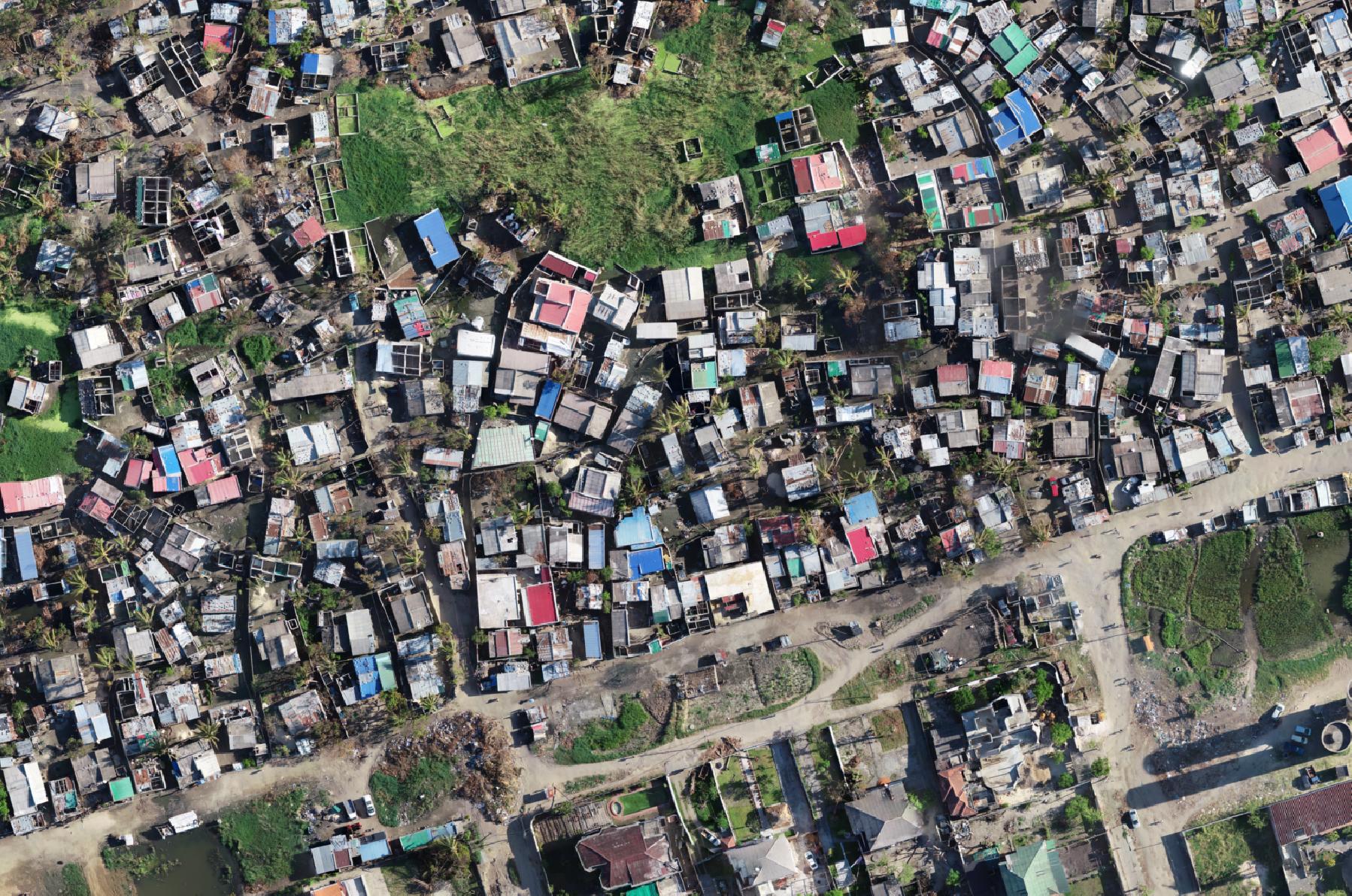}}}%
    \hspace{0.05cm}
    \subfigure[zoom-in view of the predictions]{
    {\includegraphics[width=0.32\textwidth]{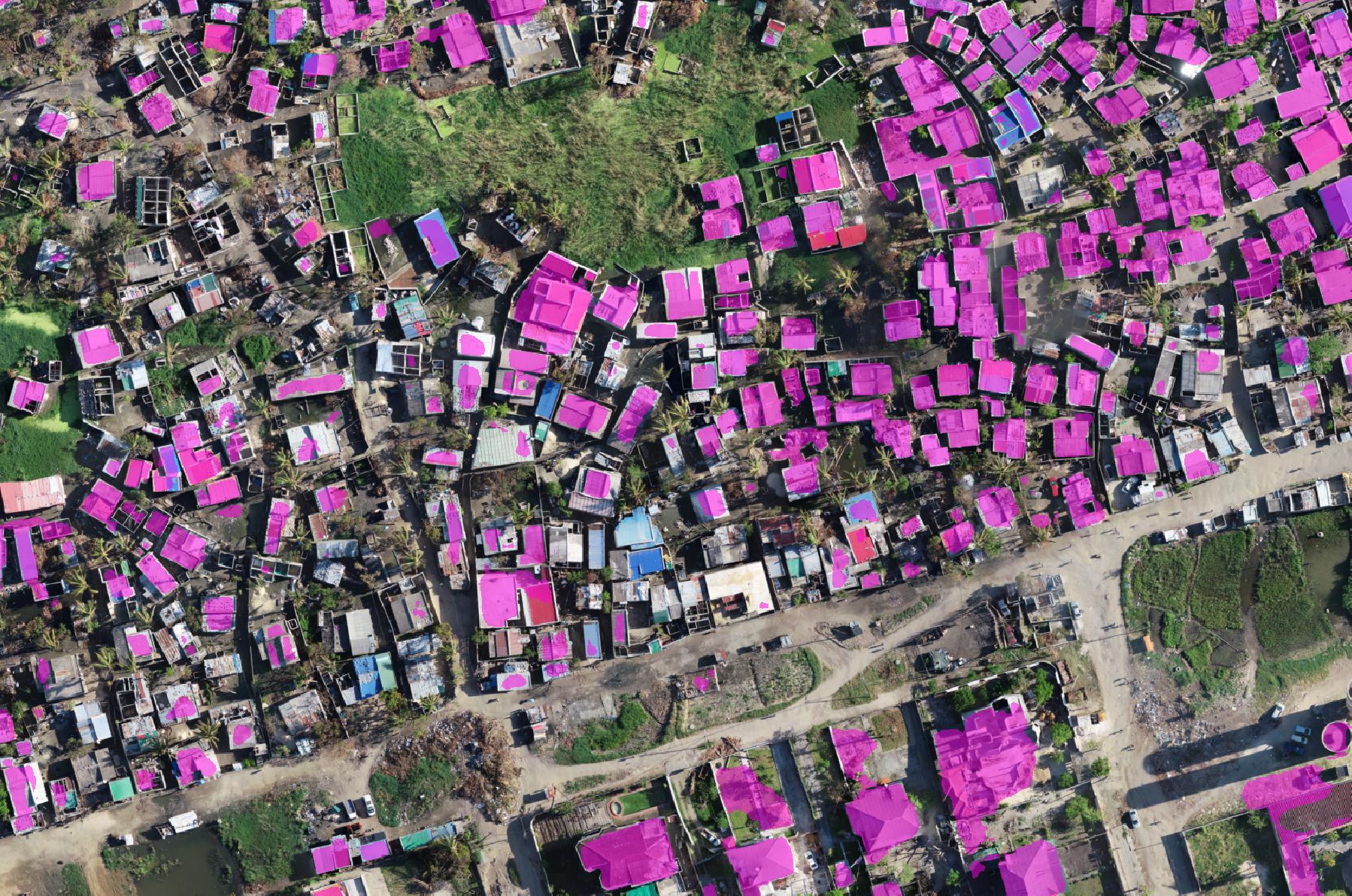}}}
    \hspace{0.05cm}
    \subfigure[zoom-in view of the ground truth]{
    {\includegraphics[width=0.32\textwidth]{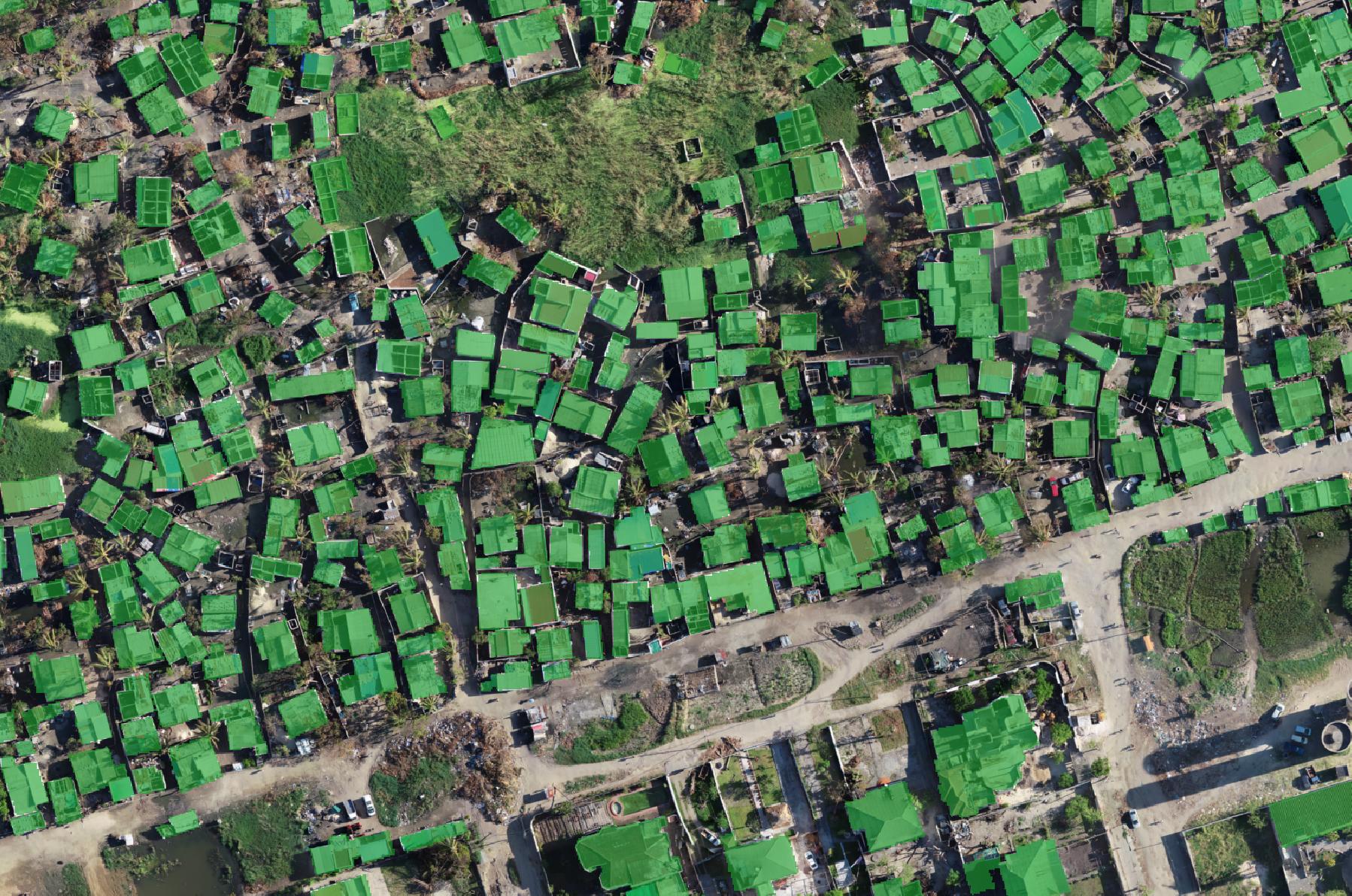}}}\\
    \vspace{-0.15cm}
    \subfigure[zoom-in view of the image]{
    {\includegraphics[width=0.32\textwidth]{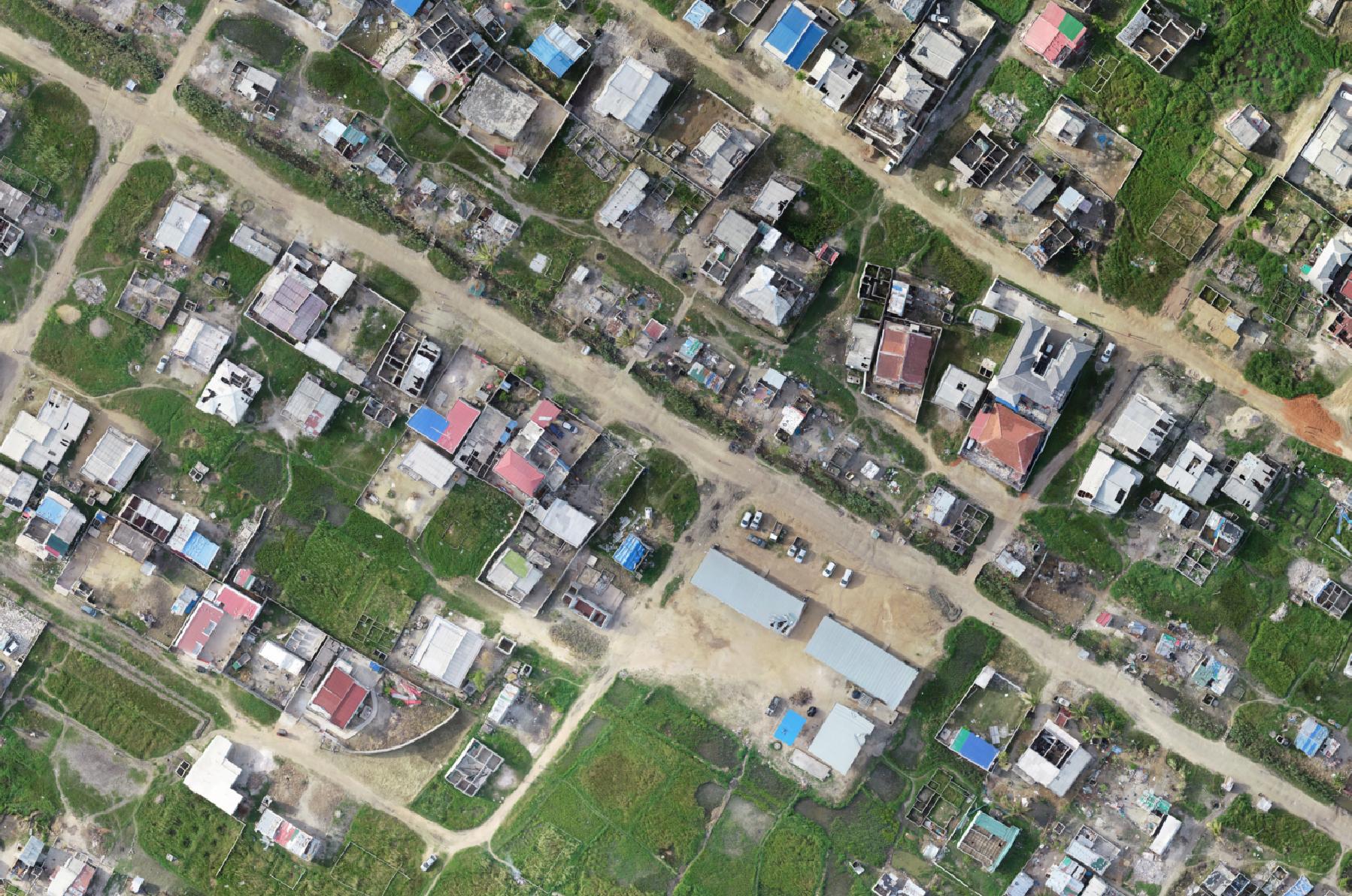}}}%
    \hspace{0.05cm}
    \subfigure[zoom-in view of the predictions]{
    {\includegraphics[width=0.32\textwidth]{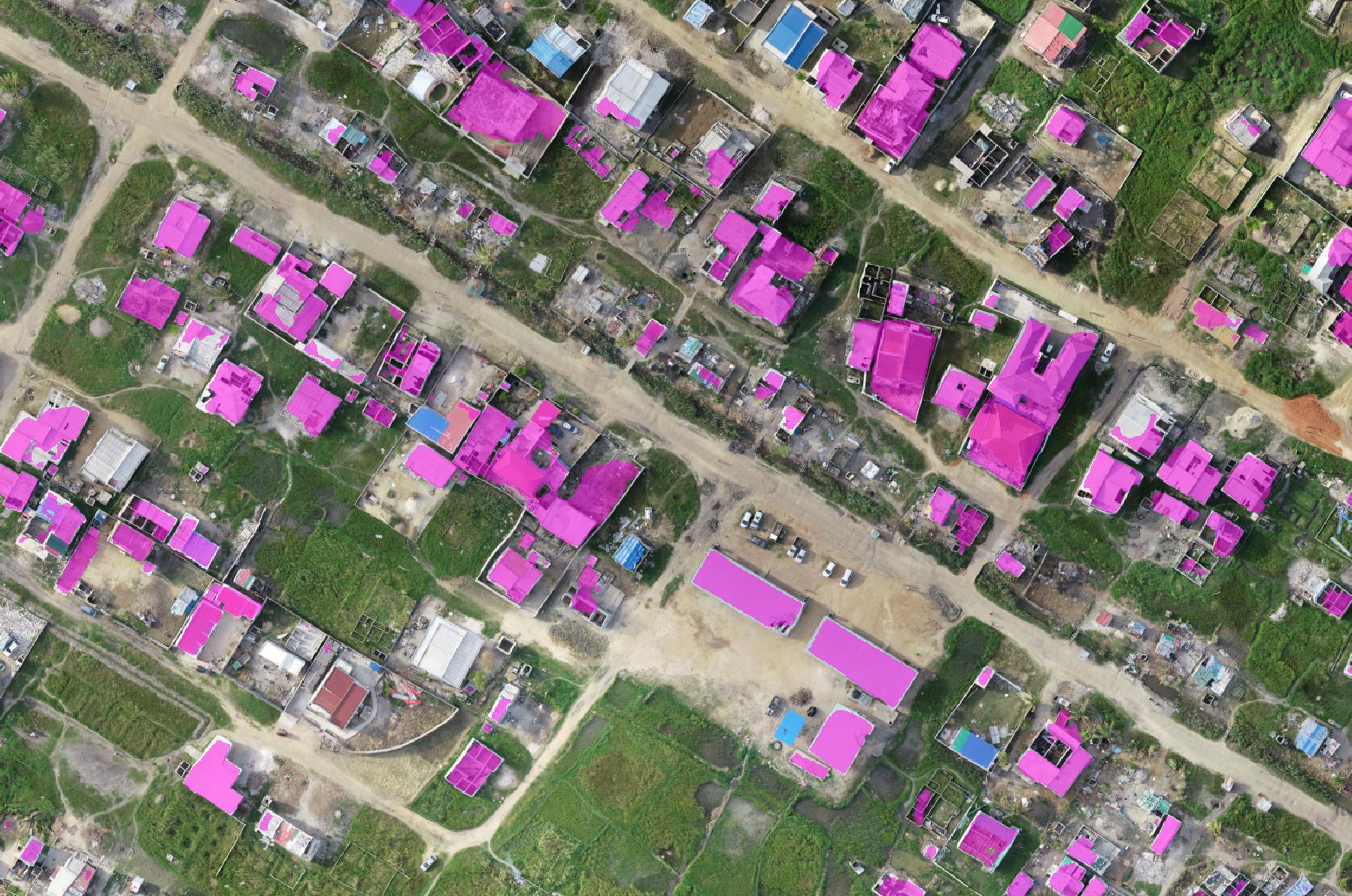}}}
    \hspace{0.05cm}
    \subfigure[zoom-in view of the ground truth]{
    {\includegraphics[width=0.32\textwidth]{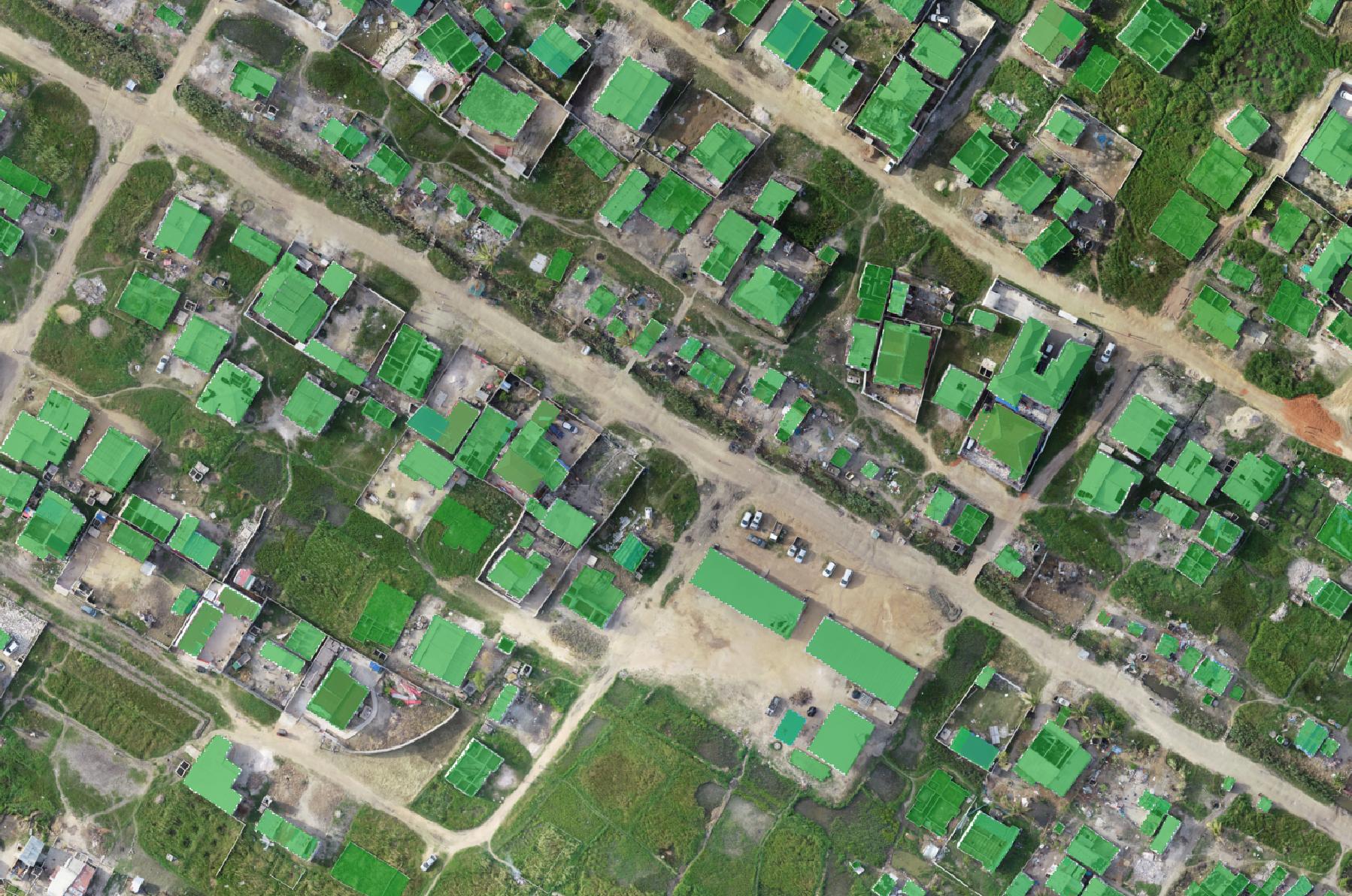}}}\\
    \vspace{-0.15cm}
    \subfigure[failure case: zoom-in view of the image]{
    {\includegraphics[width=0.32\textwidth]{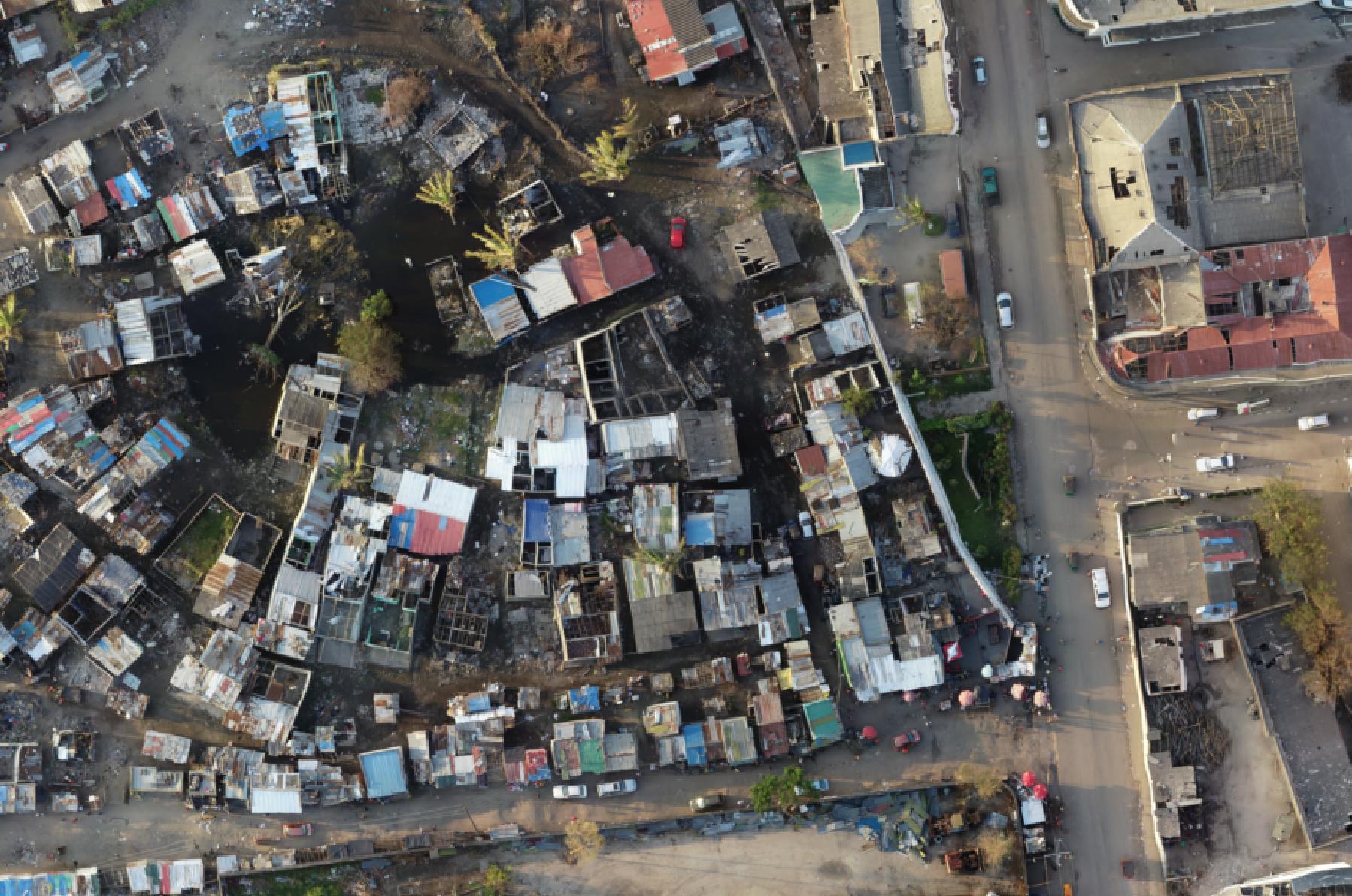}}}%
    \hspace{0.05cm}
    \subfigure[failure case: zoom-in view of the predictions]{
    {\includegraphics[width=0.32\textwidth]{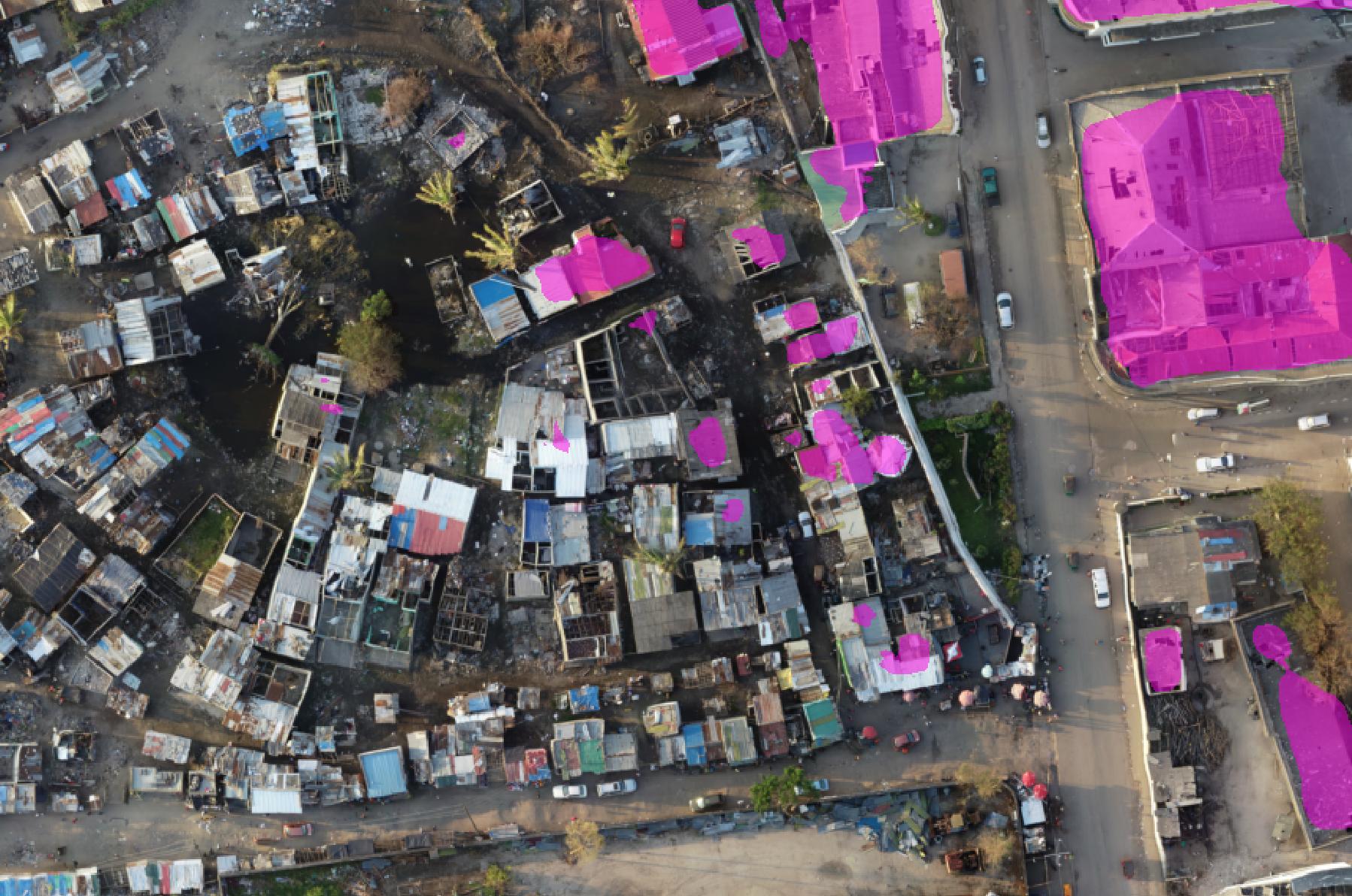}}}
    \hspace{0.05cm}
    \subfigure[failure case: zoom-in view of the ground truth]{
    {\includegraphics[width=0.32\textwidth]{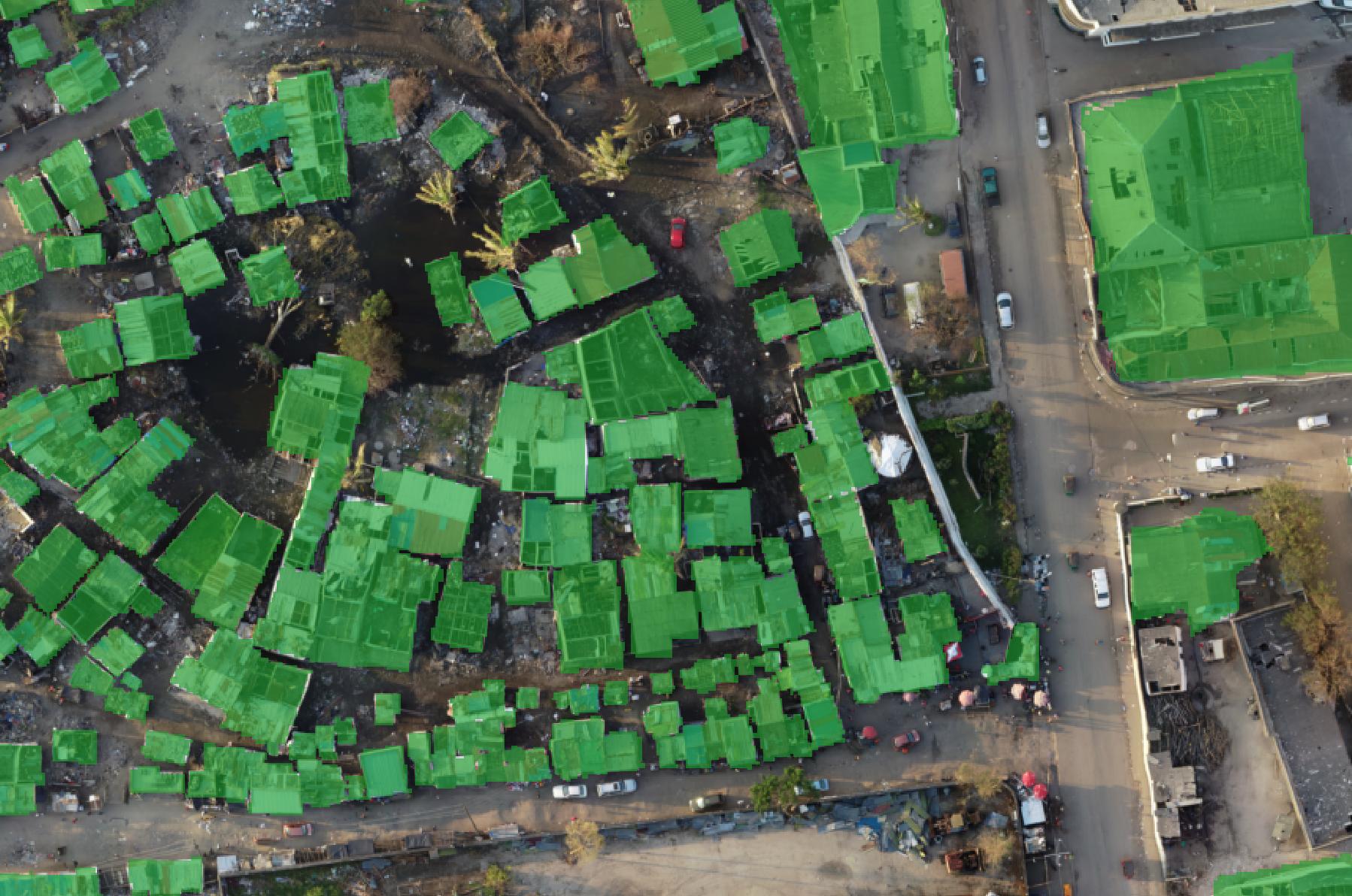}}}\\
    \caption{Building segmentation result visualization of the test area in Beira, Mozambique.}
\label{Fig:appendix_building_beira}
\end{figure*}

\textbf{Within the Beira scenes} (refer to \autoref{Fig:appendix_building_beira}), the model successfully identifies larger buildings, but fails to detect the majority of smaller structures as shown in \autoref{Fig:appendix_building_beira}~(i), which is evidenced by the precision score of 76.51\% and recall score of 44.92\%. This observation highlights the need to address domain shifts between the training and test datasets. In particular, when the building characteristics differ between the training and test areas, as in the case of Beira, factors such as different size distributions and different roof materials contribute to the observed drop in performance. Furthermore, it is also crucial to properly account for the varying imaging conditions. The Beira test data was acquired using a low-altitude UAV, resulting in a centimeter-level GSD that has unique spectral features even after downsampling. The black lines in \autoref{Fig:appendix_building_beira}~(a) resulted from co-registration with the pre-event imagery (not shown). Due to memory constraints, the UAV images are cut into smaller patches, resulting in unaligned image boundaries.

\begin{figure*}[htpb!]
\centering 
    \subfigure[Zoom-in view of the image in dense urban area ]{
    {\includegraphics[width=0.48\textwidth]{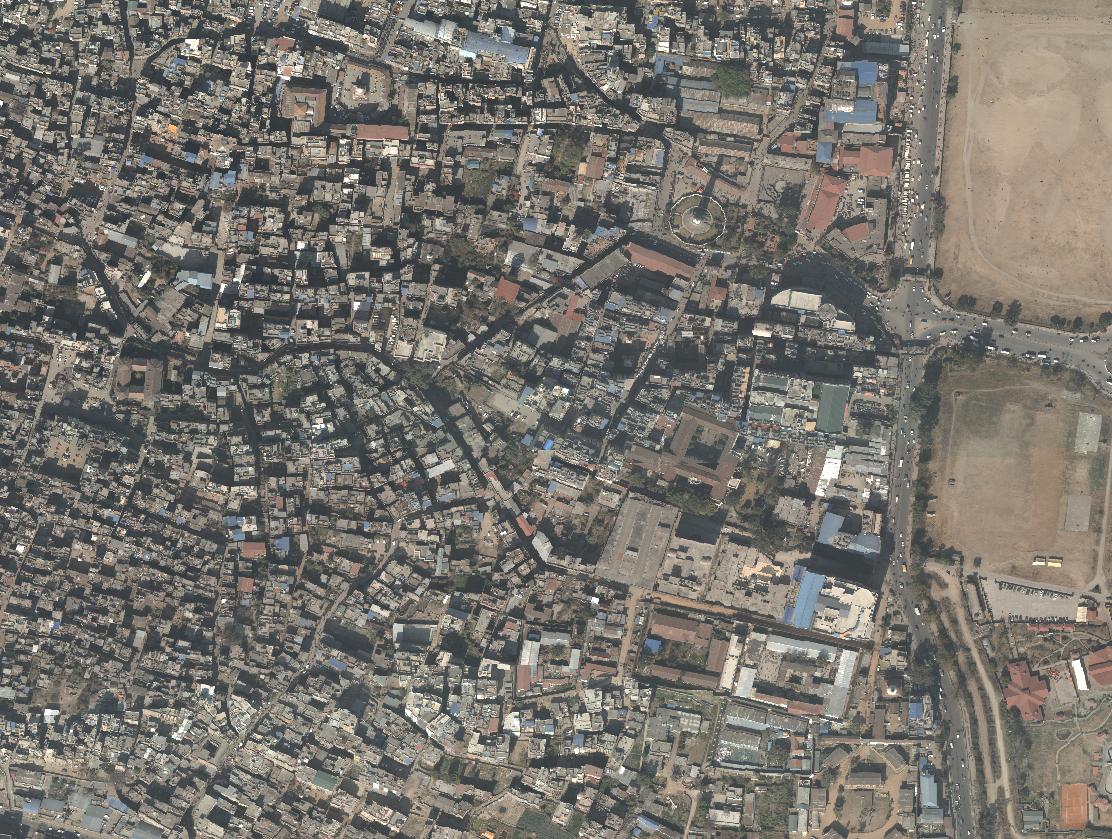}}}
    \hspace{0.05cm}
    \subfigure[Zoom-in view with the prediction in dense urban area ]{
    {\includegraphics[width=0.48\textwidth]{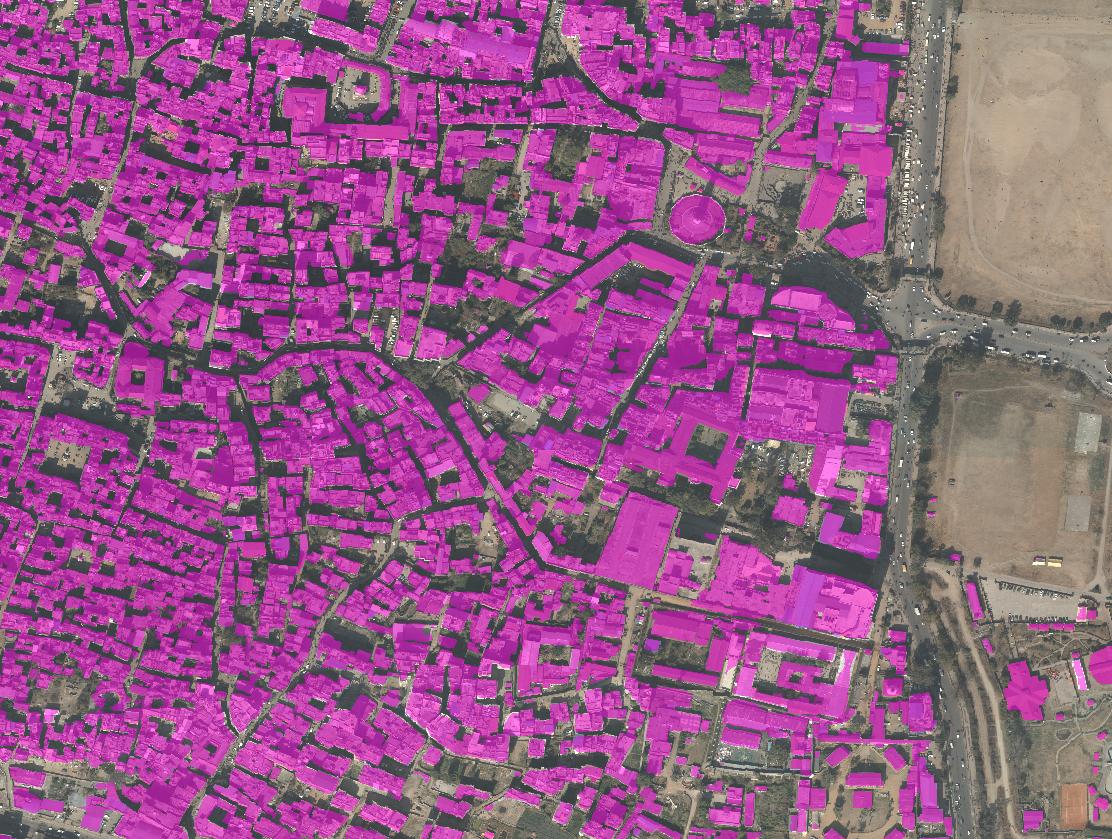}}}\\
    
    \vspace{-0.15cm}
    
    \subfigure[Zoom-in view of the image in urban area with public park  ]{
    {\includegraphics[width=0.48\textwidth]{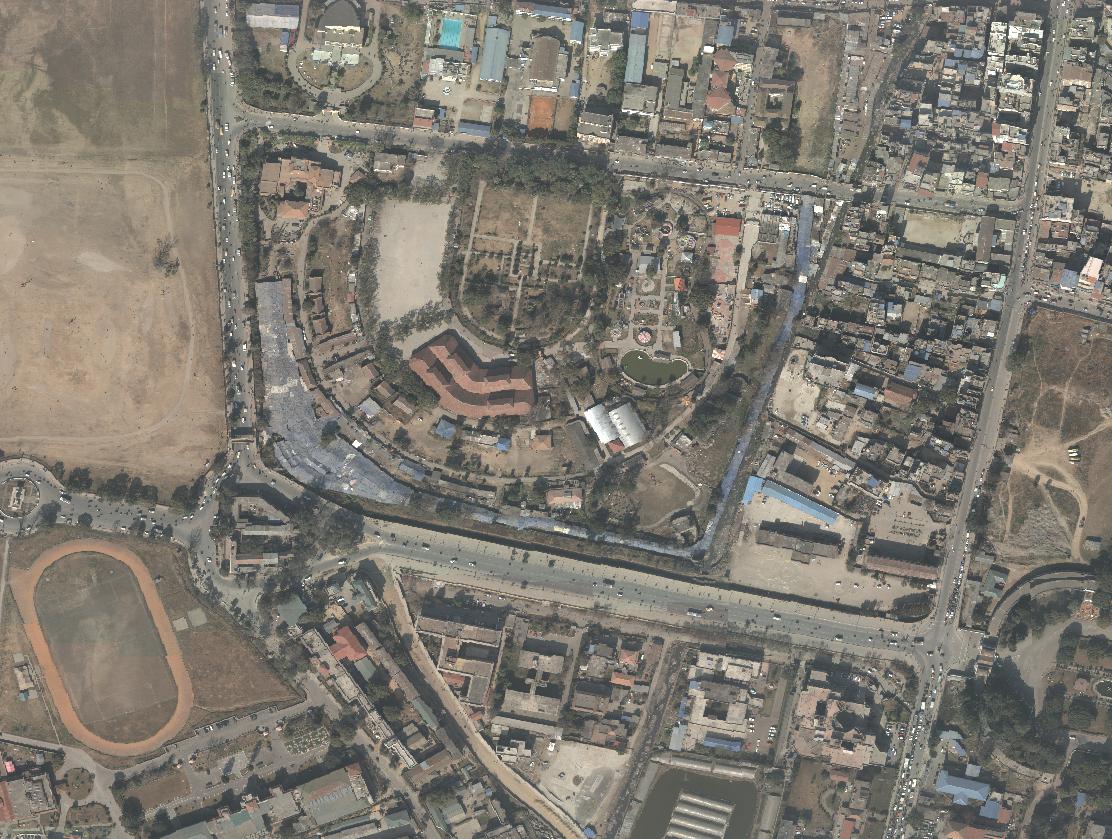}}}
    \hspace{0.05cm}
    \subfigure[Zoom-in view with the prediction in urban area with public park ]{
    {\includegraphics[width=0.48\textwidth]{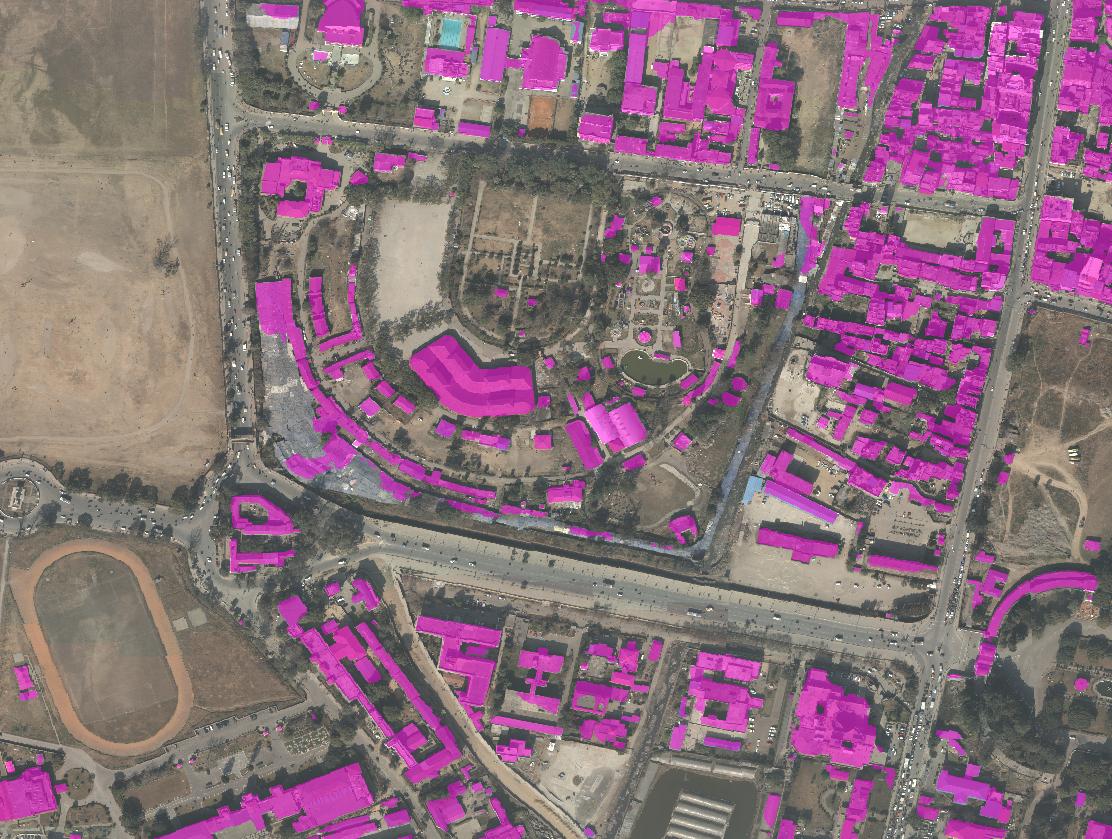}}}\\
    
    \vspace{-0.15cm}

    \subfigure[Zoom-in view of the image in urban area with more detached houses  ]{
    {\includegraphics[width=0.48\textwidth]{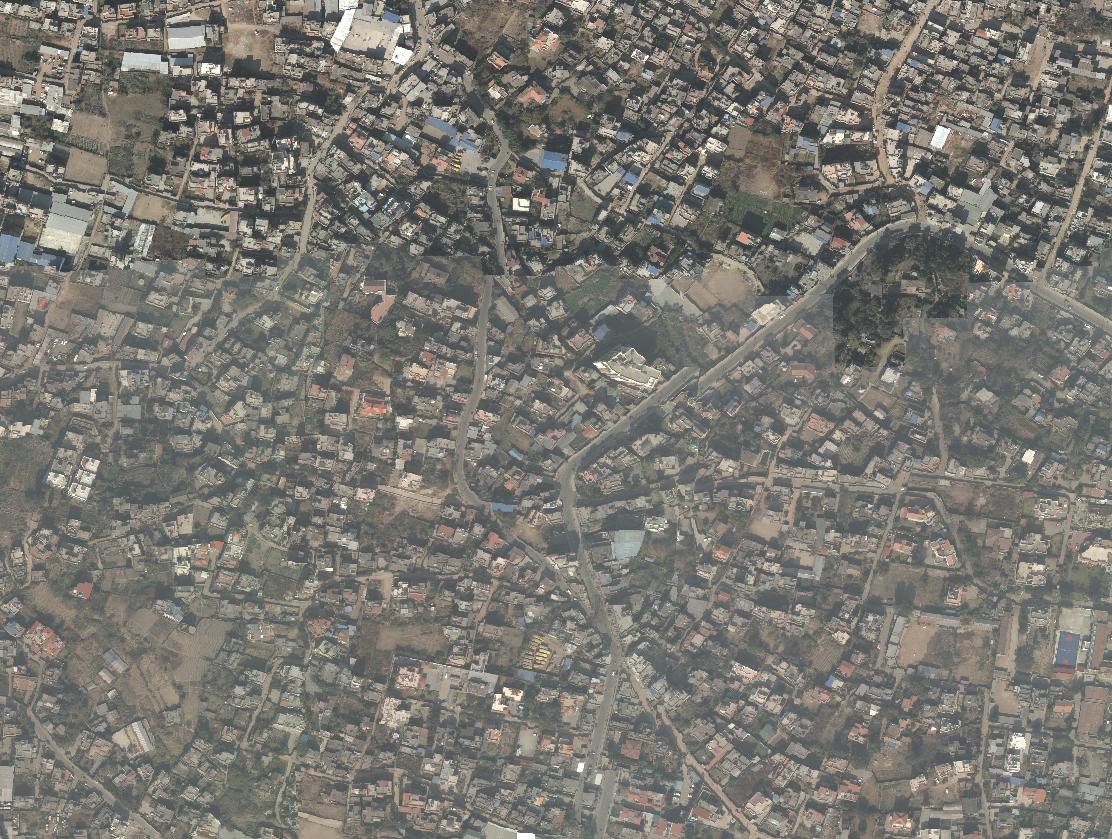}}}
    \hspace{0.05cm}
    \subfigure[Zoom-in view with the prediction in urban area with more detached houses ]{
    {\includegraphics[width=0.48\textwidth]{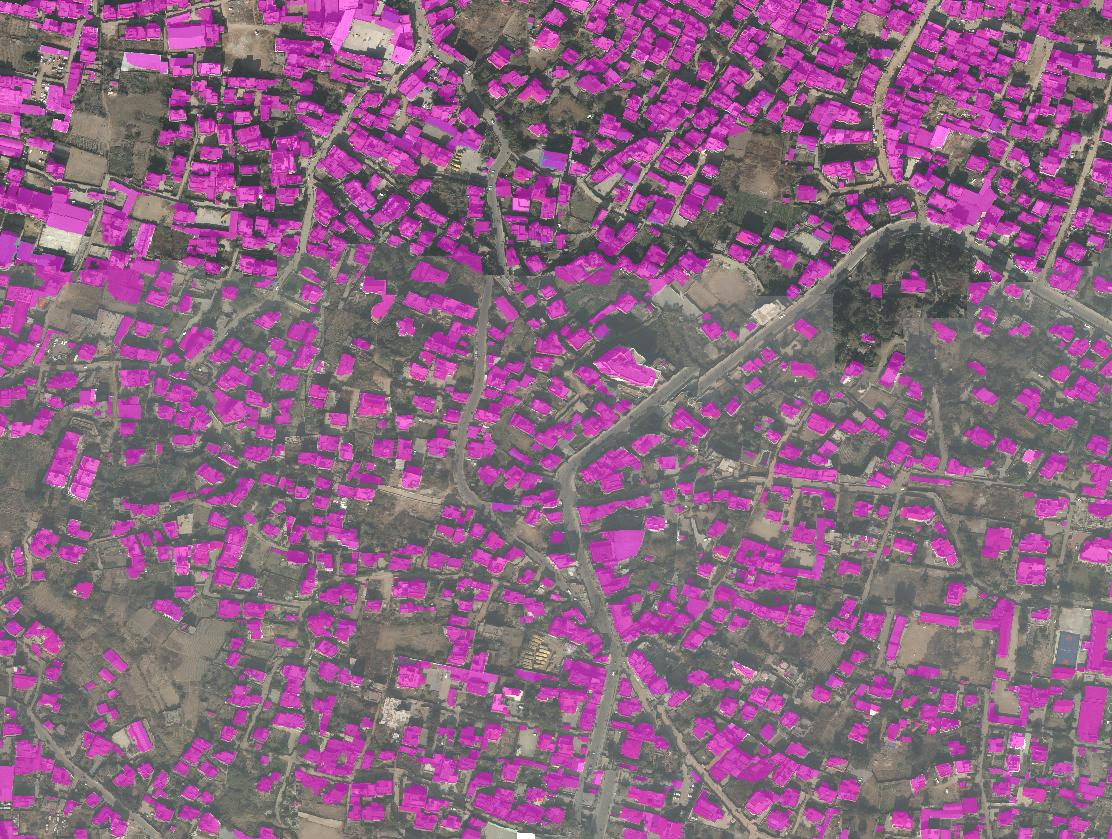}}}\\
    
    \vspace{-0.15cm}
    
    \caption{Building segmentation result visualization of the test area in Kathmandu, Nepal. Three different urban areas are selected for demonstration.}
\label{Fig:appendix_building_nepal}
\end{figure*}

In contrast to the outcomes observed in Beira, the visual outcomes achieved in \textbf{Kathmandu} (see \autoref{Fig:appendix_building_nepal}) appear promising, with the majority of buildings being successfully identified despite their dissimilarity to the training data. As ground truth data is unavailable for this scene, our analysis is conducted exclusively through qualitative evaluation. We study three different urban zones characterized by different building types and densities. \autoref{Fig:appendix_building_nepal}~(a) depicts a typical densely populated urban area with small-scale residences. Despite significant differences in building characteristics such as roof materials and density, a visual assessment shows a similar level of performance to that achieved in the European test areas. Similar results are found in the samples of \autoref{Fig:appendix_building_nepal}~(c) and \autoref{Fig:appendix_building_nepal}~(e). The success observed in the Kathmandu results underlines the robust generalizability of the method to MACS images despite the regional differences.

\subsection{Detailed results for person detection}

\begin{figure*}[t!] 
	\centering
{\includegraphics[width=1\textwidth]{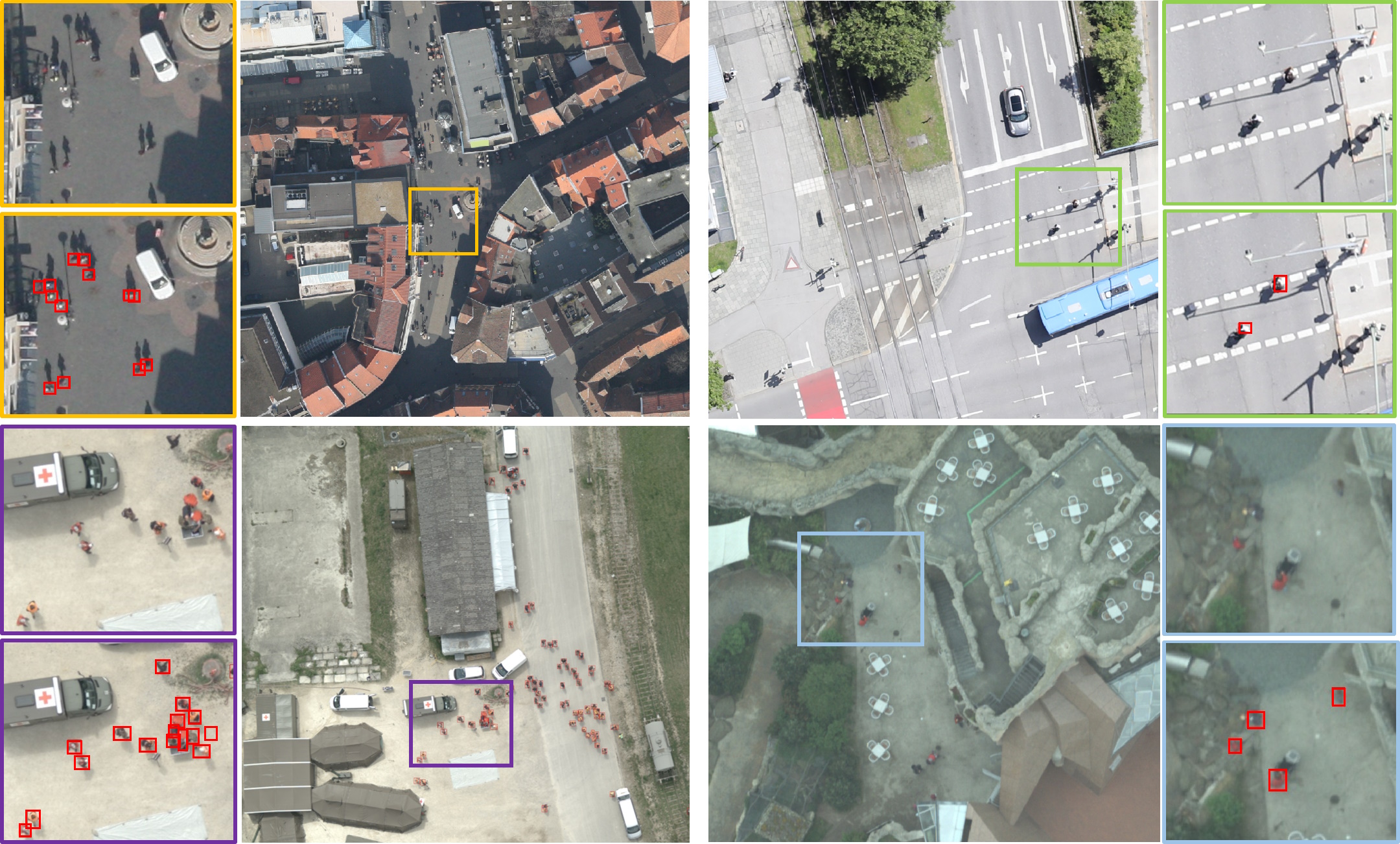}}
     \vspace{-2.5mm}
	\caption{Samples of the person detection dataset. Each person is annotated with an individual bounding box. This figure shows image samples from the training set.} 
	\label{fig:dataset_people}
\end{figure*}

\begin{figure*}[h!]
\centering 
    \subfigure[Downtown Brunswick, Germany. GSD = 3~cm, coverage = 0.1 km$^2$, process time = 99s.]{
    {\includegraphics[width=\textwidth]{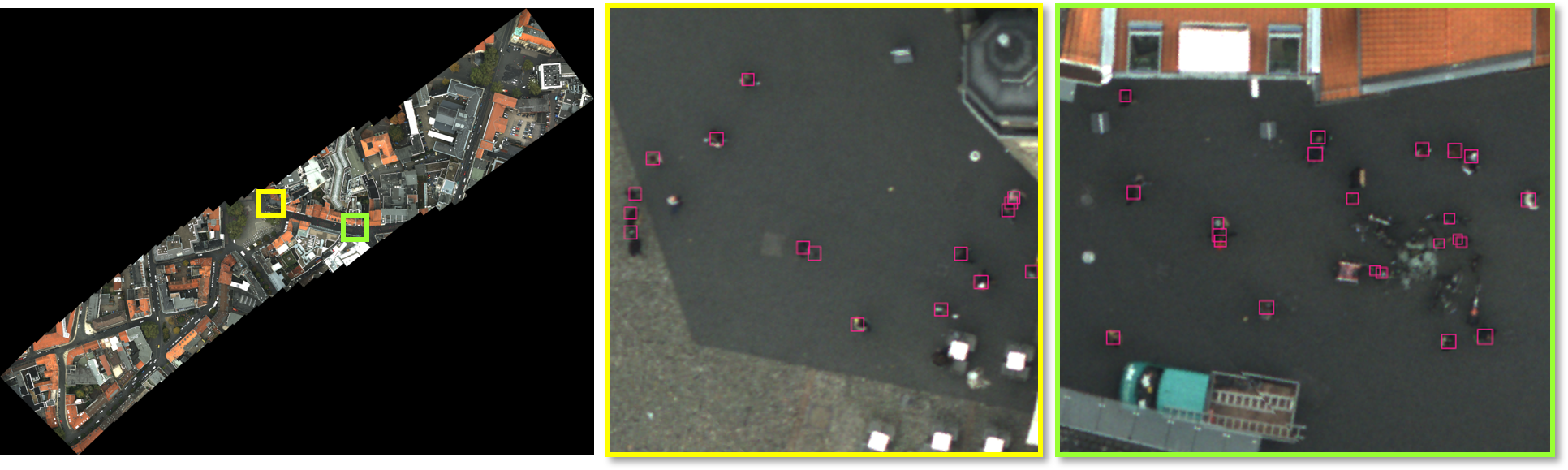}}}
    \subfigure[Epeisses, Switzerland. GSD = 3~cm, coverage = 0.08 km$^2$, process time = 47s.]{
    {\includegraphics[width=\textwidth]{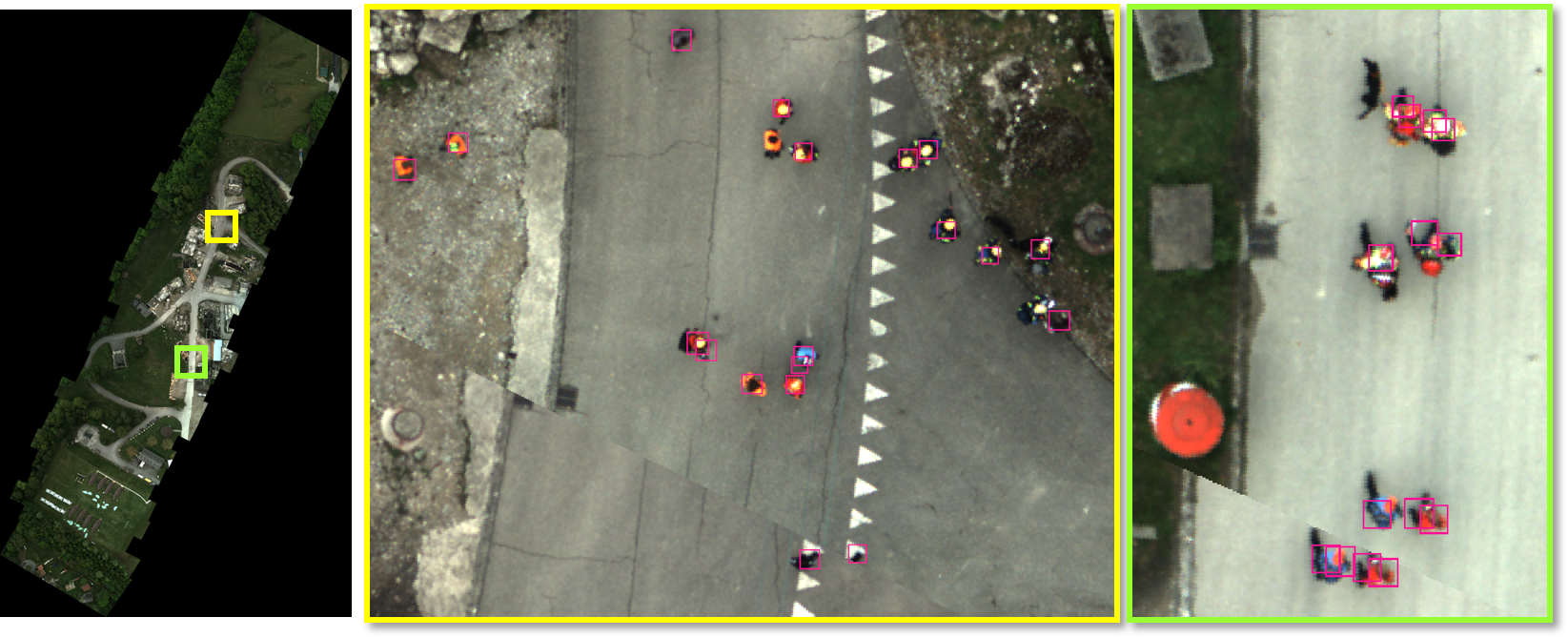}}}
    \vspace*{-0.3cm}
    \caption{Person detection on two example image mosaics processed in near real time.}
\label{Fig:appendix_person_2}
\end{figure*}

\begin{figure*}[h!]
\centering 
    \subfigure[Downtown Brunswick, Germany. GSD = 4.2~cm, AP = 67\%.]{
    {\includegraphics[width=.65\textwidth]{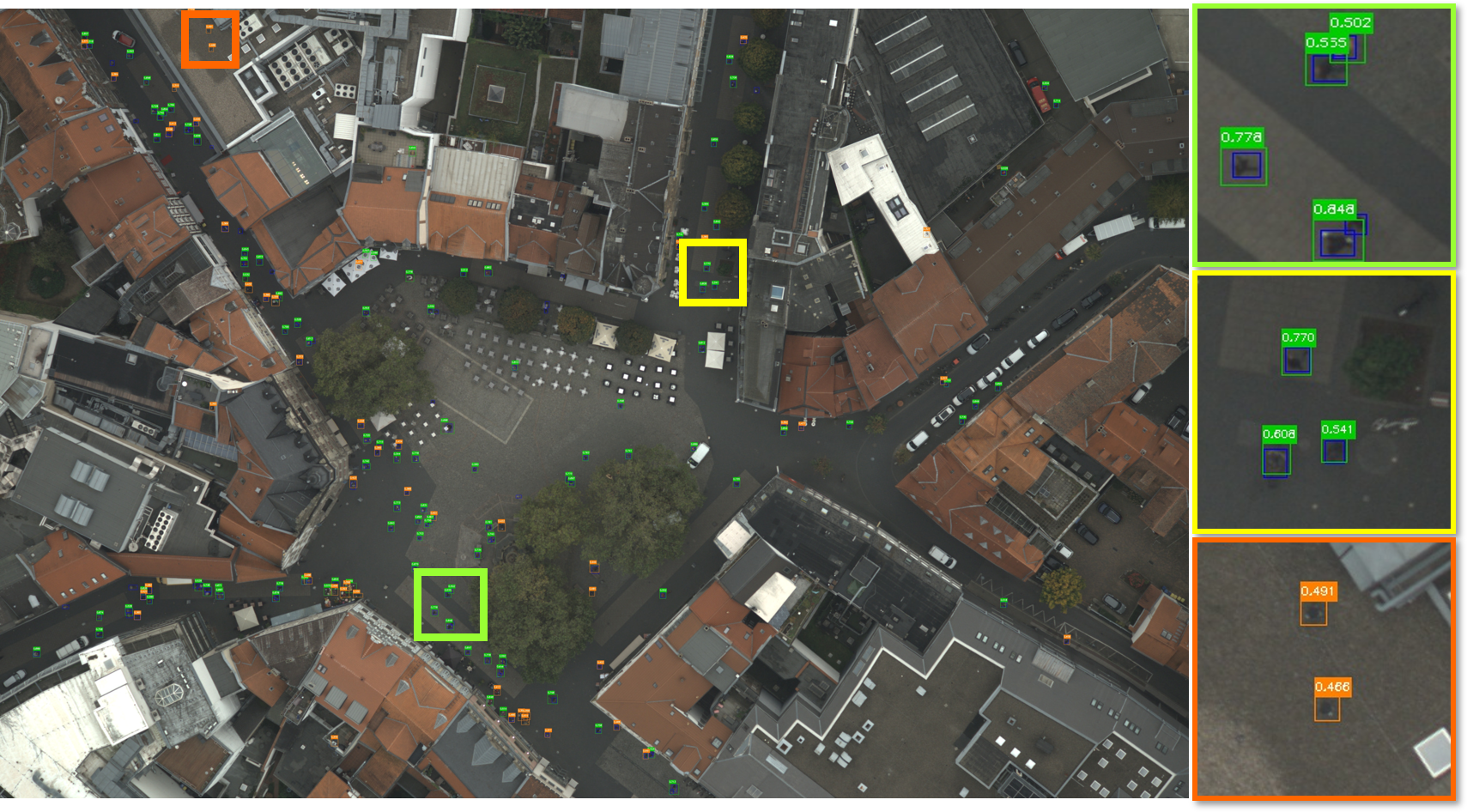}}}
    \subfigure[Flood ruins Stolberg, Germany. GSD = 1~cm, AP = 46\%.]{
    {\includegraphics[width=.65\textwidth]{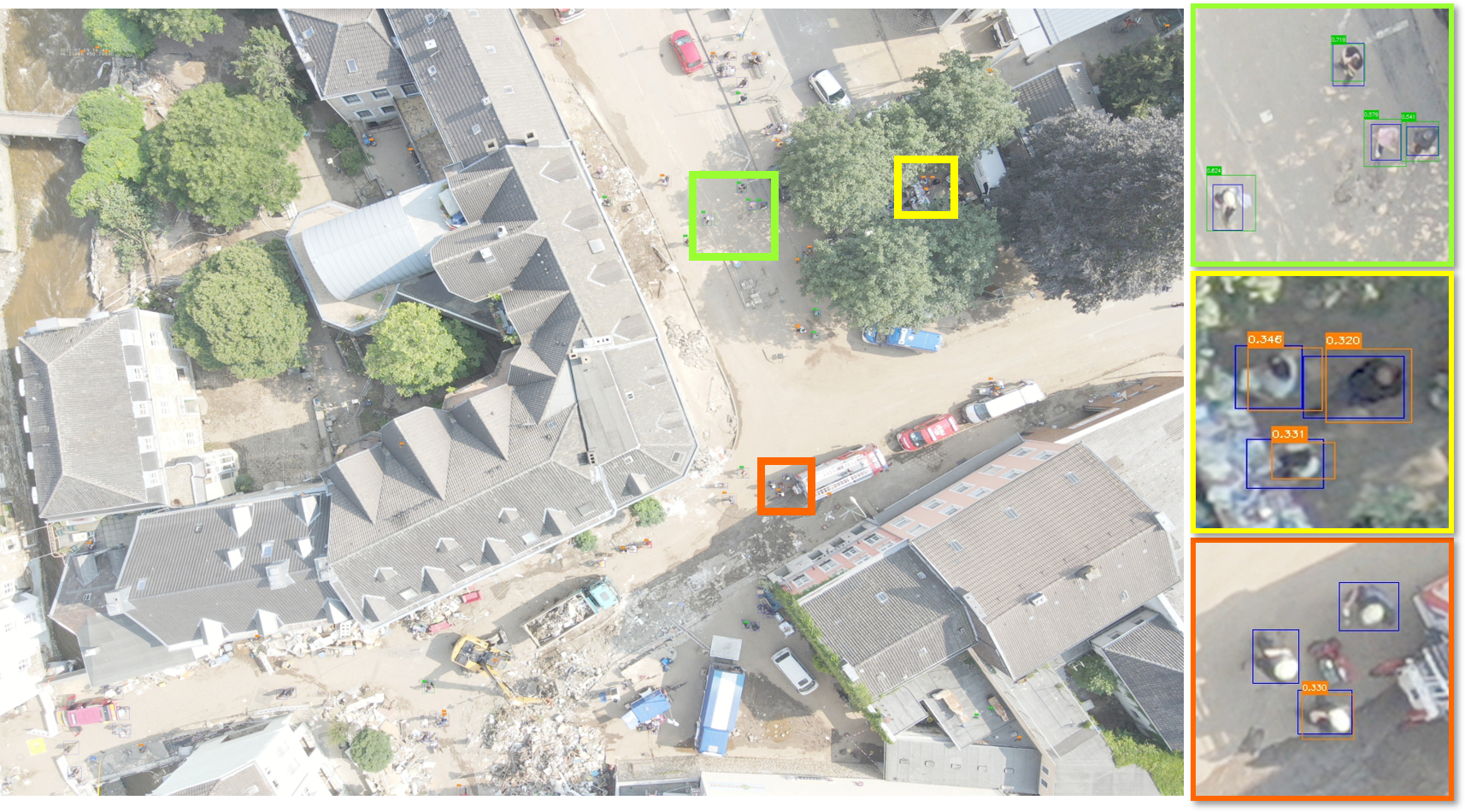}}}
    \subfigure[Villejust, France. GSD = 0.7~cm, AP = 79\%.]{
    {\includegraphics[width=.65\textwidth]{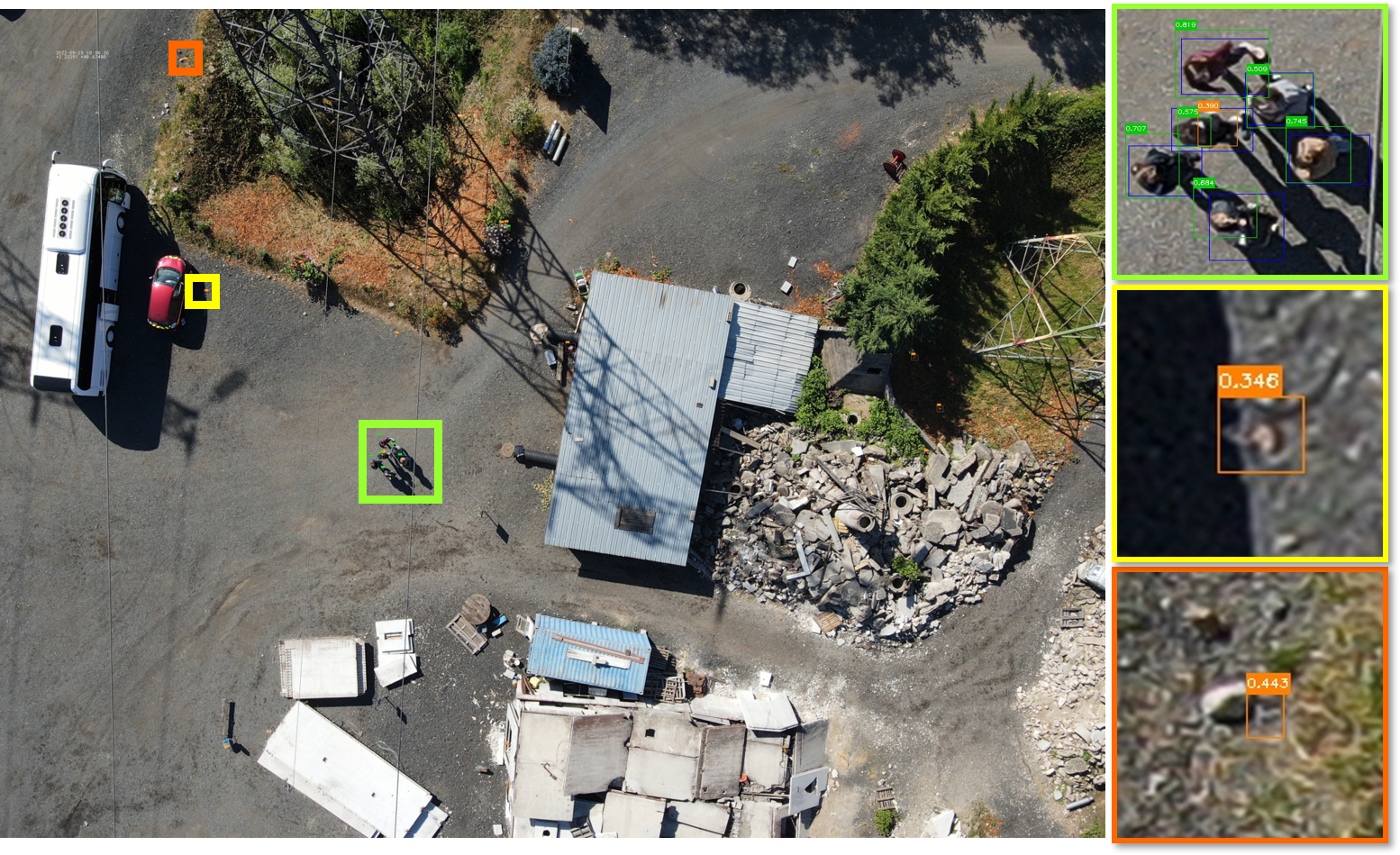}}}
    \vspace*{-0.3cm}
    \caption{Sample results from our test set for the onboard person detection with high and low confidence predictions marked in green and orange, respectively. The ground truth annotations are represented by blue bounding boxes.} 
\label{Fig:appendix_person_1}
\end{figure*}

In \autoref{fig:dataset_people}, we present selected examples from our annotated training set, where each person is individually annotated with a bounding box. Additionally, we present the results of our person detection algorithm applied to image mosaics in \autoref{Fig:appendix_person_2} for the near real time scenario and to single images for onboard processing in \autoref{Fig:appendix_person_1}. In the figures, we show zoomed areas that demonstrate both successes and failures. For example, in \autoref{Fig:appendix_person_2}~(d), the very high resolution of the image confuses the model, leading to false detections of small objects such as stones, which can resemble the appearance of people in images with larger GSDs.

\paragraph{Details of the training dataset for person detection:}
The dataset for person detection contains 311 annotated aerial and drone images acquired between 2012 and 2022 over different regions in Germany, the Netherlands, Switzerland, Spain, France, and Nepal. The sizes of the images vary between $4864\times3232$\,px, $5184\times3456$\,px, $5616\times3744$\,px, and $8000\times6000$\,px. Care was taken during the image selection process to guarantee that images with different GSD, cloud cover, and acquired with different weather conditions, sun positions, viewing angles, seasons, times of day, types of scene (urban, suburban, rural, park, and recreation sites), and application scenarios (rescue, crowd events, construction) were included in the dataset. We divided the $311$ images of our dataset into three disjoint sets: 1) the training set consisting of $259$ images with $6934$ annotations, 2) the validation set consisting of $25$ images with $2706$ annotations, and 3) the test set consisting of $27$ images with $410$ annotations. The samples in \autoref{fig:dataset_people} illustrate the diversity of the images within the dataset.

\end{appendices}

\end{document}